\documentstyle[12pt,amsmath,amssymb,amsfonts,amsthm,amssymb,amscd,amsbsy,amstext,epsfig]{article}

\numberwithin{equation}{section}

\def\ben{\begin{equation}}
\def\een{\end{equation}}
\def\half{{\textstyle{1\over2}}}
\let\a=\alpha   \let\d=\delta 
  \let\q=\theta 
\let\l=\lambda     \let\r=\rho

\def\be{\begin{equation}}
\def\ee{\end{equation}}
\def\ba{\begin{array}}
\def\ea{\end{array}}

\newcommand{\bea}{\begin{eqnarray}}
\newcommand{\eea}{\end{eqnarray}}

\def\BS{\boldsymbol{S}}
\def\Bphi{\boldsymbol{\phi}}
\def\BPhi{\boldsymbol{\Phi}}
\def\Bx{\boldsymbol{x}}

\def\N{{{\Bbb N}}}

\def\xb{{\boldsymbol x}}
\def\vx{\vec{x}}
\def\vy{\vec{y}}
\def\vz{\vec{z}}

\def\lab{\label}
\def\6{\partial}

\def\le{\left}
\def\ri{\right}

\def\N{{\cal N}}

\def\Dbarslash{\,\,{\raise.15ex\hbox{/}\mkern-12mu {\bar D}}}
\def\Dslash{\,\,{\raise.15ex\hbox{/}\mkern-12mu D}}
\def\delslash{\,\,{\raise.15ex\hbox{/}\mkern-9mu \partial}}
\def\delbarslash{\,\,{\raise.15ex\hbox{/}\mkern-9mu {\bar\partial}}}

\def\BOmega{\boldsymbol{\Omega}}

\newcommand{\EQ}[1]{\begin{equation} #1 \end{equation}}

\newcommand{\SP}[1]{\begin{equation}\begin{split} #1
\end{split}\end{equation}}

\begin{document}
\begin{flushright}
CPHT-RR012.0307, LPTENS 07/11, NSF-KITP-07-26 \\
hep-th/0703100
\end{flushright}

\begin{center}
\vspace{0.2cm} { \LARGE {\bf Topology change in commuting saddles of \\
\vspace{0.2cm}
thermal ${\mathcal{N}}=4$ SYM theory}}

\vspace{0.5cm}

Umut G\"ursoy${}^1$, Sean A. Hartnoll${}^2$, Timothy J. Hollowood${}^3$ and S. Prem Kumar${}^3$

\vspace{0.4cm}

{\it ${}^1$ CPHT, \'Ecole Polytechnique, UMR du CNRS 7644,
            91128 Palaiseau, France} \\
and\\
{\it LPT, \'Ecole Normale Sup\'erieure, 75231 Paris 05, France}
            {\tt gursoy@cpht.polytechnique.fr } \\

\vspace{0.3cm}

{\it  ${}^2$ KITP, University of California Santa Barbara \\
 CA 93106, USA } \\
 {\tt hartnoll@kitp.ucsb.edu} \\

\vspace{0.3cm}

{\it ${}^3$ Department of Physics, University of Wales Swansea \\
        Swansea, SA2 8PP, UK } \\
{\tt s.p.kumar@swansea.ac.uk,  t.hollowood@swansea.ac.uk}

\vspace{0.3cm}

\end{center}

\begin{abstract}

We study the large $N$ saddle points of weakly coupled
${\mathcal{N}}=4$ super Yang-Mills theory on $\BS^1\times \BS^3$
that are described by a commuting matrix model for the seven
scalar fields $\{ A_0,
\Phi_J\}$. We show that at temperatures below the
Hagedorn/`deconfinement' transition the joint eigenvalue
distribution is $\BS^1
\times \BS^5$. At high temperatures $T\gg 1/R_{\BS^3}$, the
eigenvalues form an ellipsoid with topology $\BS^6$. We show how
the deconfinement transition realises the topology change $\BS^1
\times \BS^5 \to \BS^6$. Furthermore, we find compelling evidence
that when the temperature is increased to $T = 1/(\sqrt\lambda
R_{\BS^3})$ the saddle with $\BS^6$ topology changes continuously
to one with $\BS^5$ topology in a new second order quantum phase
transition occurring in these saddles.
\end{abstract}

\pagebreak
\setcounter{page}{1}

\tableofcontents

\pagebreak

\section{Introduction and summary}

In this work we will combine two
strands of research that have each generated interesting results
within the AdS/CFT correspondence \cite{Maldacena:1997re} over the
past couple of years.

On the one hand, it has been understood how the dynamics of
certain supersymmetric sectors of ${\mathcal N} = 4$ super
Yang-Mills theory on a spatial $\BS^3$ are described by matrix
models for the scalar fields $\Phi_J$ of the theory. Furthermore,
it was discovered that the eigenvalue distributions of these
scalars directly reconstruct a dual spacetime geometry
\cite{Berenstein:2004kk, Lin:2004nb, Berenstein:2005aa,
Berenstein:2007wz}. These works have provided a striking
realisation of emergent geometry within the AdS/CFT setup.

On the other hand, in the same theory at finite temperature it was
shown that at weak 't Hooft coupling there is a phase transition
at a critical temperature \cite{Sundborg:1999ue,Aharony:2003sx}.
This transition appears to be similar to the Hawking-Page
transition at strong coupling, which describes the appearance of a
black hole in the dual geometry \cite{Hawking:1982dh,
Witten:1998zw}. In particular, for both phase transitions the
eigenvalue distribution of the time component of the gauge field,
$A_0$, becomes a nonuniform distribution on $\BS^1$ at the
transition and ultimately becomes a gapped distribution as the
temperature is increased.

Combining the insights of these works, one might hope that
studying the joint eigenvalue distribution of $\{A_0,\Phi_J\}$
could lead to an understanding of the dual black hole geometry in
the weak coupling regime. In fact, a fundamental question is
whether or not there is a well defined sense in which it is useful
to conceive of a weakly coupled plasma as being dual to a black
hole. Various recent works have posed this as a question about
correlators in real time physics \cite{Fidkowski:2003nf,
Festuccia:2005pi, Hartnoll:2005ju, Hubeny:2006yu,
Festuccia:2006sa}. The question is not purely of conceptual
interest, as some discussion of the fireball created in the
Relativistic Heavy Ion Collider attests, see for instance
\cite{Kovtun:2004de, Nastase:2005rp}. The approach here
will be Euclidean and hence concerned with equilibrium physics.

In most works on the finite temperature theory at weak coupling,
the scalar fields are integrated out on the grounds that they
acquire thermal masses, as well as having a classical mass due to
being conformally coupled to a spatial $\BS^3$. However, at
temperatures $TR \sim \lambda^{-1/2}$, the one loop thermal mass
of $A_0$ is of the same magnitude as that for the scalar fields,
so $A_0$ and $\Phi_J$ (and indeed the $A_i$ \cite{Aharony:2007rj})
should participate equally in the dynamics. The full dynamics of
the theory is difficult to study, as there are infinitely many
coupled modes that can condense, and furthermore generic
condensates will not commute. In this work we will make the
drastic (but consistent) simplification of only considering
saddles in which the homogeneous modes of $\{A_0,\Phi_J\}$
condense and where the condensates commute. We believe that these
are interesting saddles to consider; they are the most `geometric'
of possible saddles at weak coupling and can be described very
explicitly. They may also have a connection with the dominant
geometrical saddles at strong coupling. However, it is clear
\cite{Aharony:2007rj} that they are not the absolute minima of the
theory. Phase transitions in these saddles are not transitions of
the full theory.

The effective potential necessary for this study, the potential
for commuting scalars and $A_0$, was computed in
\cite{Hollowood:2006xb}, who used it to show that the potential
reveals a weak coupling analogue of the Gregory-Laflamme
instability of the small black hole towards localisation on
$\BS^5$ \cite{Hollowood:2006xb}. In
\cite{Hartnoll:2006pj} it was found that allowing one scalar field
to be non-zero leads to a non-trivial joint eigenvalue
distribution of $\{A_0,\Phi_J\}$. This is due to a logarithmic
repulsion between the scalar eigenvalues at large $N$ overcoming
the classical and thermal mass terms. However, the solutions found
in \cite{Hartnoll:2006pj} break R symmetry in picking out a given
scalar field.

In this work we find the eigenvalue distribution of
$\{A_0,\Phi_J\}$ that minimises the one loop effective potential
of ${\mathcal N} = 4$ super Yang-Mills theory on $\BS^1 \times
\BS^3$, when restricted to commuting matrices for the homogeneous modes
of $A_0$ and $\Phi_J$. Our results cover the whole range of
temperatures and couplings for which the one loop potential is
valid: $TR \ll\lambda^{-1}$. We use $R$ to denote the radius of
the spatial $\BS^3$. As well as finding the minimal action
distribution, which preserves $SO(6)$ R symmetry, in Appendix
\ref{sec:saddles} we exhibit other interesting saddle points of
the effective action with various patterns of R symmetry breaking.

In the supersymmetric sector, the logarithmic repulsion between---and
subsequent condensation of---scalar eigenvalues is crucial for
the emergence of a dual spacetime geometry in the large $N$ limit
\cite{Berenstein:2005aa, Berenstein:2007wz}.
This effect needs to be considered also at finite temperature. A
central result of our work is that at high temperatures
$TR=\lambda^{-1/2}$, the condensate of scalar eigenvalues
backreacts sufficiently onto the distribution of the $A_0$
eigenvalues to cause a new phase transition, in the commuting
matrix saddle.

\subsection{Summary of results on the eigenvalue distribution}

At low temperatures we find analytically that the eigenvalues are
uniformly distributed as $\BS^1\times \BS^5$. Here the $\BS^1$ is
in the $A_0$ direction whereas the $\BS^5$ is in the $\Phi_J$
directions and has a radius that scales like $\lambda/R$.

At the deconfinement temperature, $TR_c = -1/\log(7-4\sqrt{3})
\approx 0.38$ \cite{Aharony:2003sx}, the distribution on $\BS^1$
develops a gap. We show analytically that this implies that the
full distribution acquires topology $\BS^6$.

At high temperatures, $1 \ll TR \ll \lambda^{-1} $, we find two
candidate saddle points. There is an $\BS^5$ distribution that can
be described analytically and an ellipsoidal distribution with
topology $\BS^6$ that we obtain analytically for $TR \ll
\lambda^{-1/2}$ and numerically at higher temperatures. The
$\BS^5$ solution is at a point in the $A_0$ direction. Both
solutions preserve the full $SO(6)_R$ symmetry.

At weak coupling, we find that the ellipsoidal solution has lowest
action at temperatures $TR_c\leq TR<\lambda^{-1/2}$. The first
result of this paper is therefore that, in the sector of commuting
spatially homogeneous fields, the low and high temperature phases
of weakly coupled ${\mathcal{N}}=4$ SYM theory on $\BS^3$ are
characterised by eigenvalue distributions with differing topology:
$\BS^1 \times
\BS^5$ versus $\BS^6$. As the temperature is increased to the
value $TR=\lambda^{-1/2}$ at fixed weak coupling, or equivalently,
as the coupling is increased up to $\lambda =1/(TR)^2$ at fixed
high temperatures, we find compelling evidence for a second order
phase transition. In this transition the $\BS^6$ solution smoothly
collapses to $\BS^5$. We summarize the behaviour of these saddles
in figure 1 below. Details will be discussed below.
\begin{figure}[h]
\begin{center}
\epsfig{file=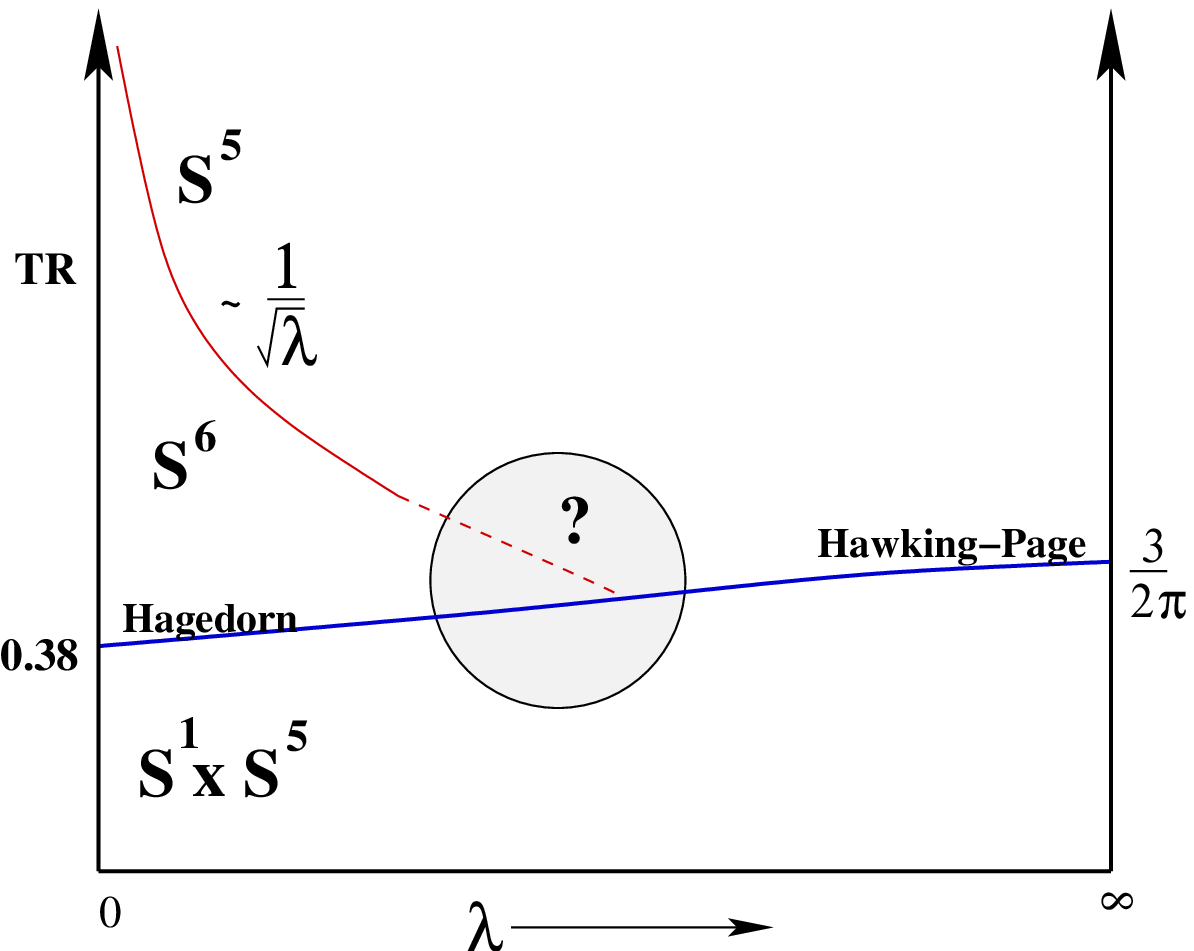,width=3.5in}
\end{center}
\noindent \small {\bf Figure 1:} Topology of the dominant eigenvalue distribution
in which the only the homogeneous modes of $\{A_0,\Phi_J\}$
condense and furthermore commute.
\label{phase1}
\end{figure}

\section{The effective potential}

Our starting point is the one loop effective potential of
${\mathcal{N}}=4$ SYM theory on $\BS^1\times \BS^3$, restricted to
a certain sector of the theory. We denote the radius of the
$\BS^3$ by $R$ and $\beta=1/T$ will be the circumference of the
thermal $\BS^1$. The effective potential is the result of
integrating out all inhomogeneous modes on the three sphere and
depends only on the homogeneous modes. The latter are the vacuum
expectation values of the six scalar fields in the theory,
$\Phi_J$, $J=1,2,\ldots ,6$,\footnote{We denote a six vector with
a boldface symbol, {\it e.g.\/}~$\BPhi$ and $\Bphi$.} and the time
component of the gauge field, $A_0$. This potential was computed
in \cite{Hollowood:2006xb} and also discussed recently in
\cite{Hartnoll:2006pj}.

An important feature of the effective potential is that it is
evaluated on the space of mutually commuting scalar homogeneous modes.
The potential is therefore a function of
the eigenvalues of the adjoint scalars,
which we denote by $\{\Bphi_{p}\}$, $p=1,2,\ldots ,N$,
and the eigenvalues of $\beta A_0$, which we denote by $\{\theta_p\}$.

It is natural to consider the effective potential on the space of
commuting scalar field expectation values. In flat space and at
zero temperature, these characterize the moduli space of vacua of
the ${\cal N}=4$ theory, wherein off diagonal modes obtain masses
$|{\boldsymbol\phi}_{pq}|$.\footnote{We set $|\Bphi|\equiv
\sqrt{\sum_J\phi_J^2}$ and
  $\Bphi_{pq}=\Bphi_p-\Bphi_q$.}
There is no moduli space on $\BS^3$ due to the classical conformal
coupling to the background scalar curvature. However, around a
homogeneous background with commuting VEVs for the scalars, off
diagonal fluctuations of the scalar zero modes obtain a positive
mass given by $\sqrt{R^{-2}+|{\boldsymbol\phi}_{pq}|^2}$
and those of $A_0$ have masses $|{\boldsymbol\phi}_{pq}|$.
For this reason all the configurations that we consider will be
locally stable against small fluctuations involving the
off-diagonal entries of the scalar field matrices. This is also
manifest in the reality of the one loop effective potential below,
on the space of simultaneously diagonal configurations.

We should emphasize that whereas the truncation to commuting
matrices is consistent, it will not describe the absolute minimum
of the action \cite{Aharony:2007rj}. We will see however, that
these commuting saddles display some interesting dynamics.
Throughout this work, we are only considering saddles in which the
matrices commute.

A second consistent truncation we have made is to consider
condensates for only the homogeneous modes. At infinite $N$, the
inhomogeneous modes can also condense without breaking the spatial
$SO(4)$ rotational invariance. The full dynamics should include
these modes also.

There are three contributions to the potential: the classical
conformal mass term for the scalars and then two determinants
arising from integrating out the bosons and fermions in the theory
\be
S_{\rm eff}[\Bphi_{p}, \theta_p]\;=\; S^{(0)}+S^{(1)}_{\rm b} +
  S^{(1)}_{\rm f} \,. \label{full}
\ee
The classical term is
\be
S^{(0)}=\frac{\beta R\pi^2N}{\lambda}\sum_{p=1}^{N}
|\Bphi_{p}|^2\,.
\ee
Integrating out the gauge fields, scalar and ghost fluctuations
yields the bosonic contribution
\SP{
S_{\rm b}^{(1)}&= \sum_{pq=1}^{N}\Big\{-\log
\left|\sinh \frac{\beta |{\boldsymbol{\phi}}_{pq}|+i
    \theta_{pq}}2\right|-\log2
\\ &\qquad+\sum_{\ell=0}^{\infty} 2 (2 \ell+3)(2\ell+1)
\Big(\frac\beta2\sqrt{(\ell+1)^2R^{-2}+|\Bphi_{pq}|^2}\\ &\qquad\qquad+
\log
\Big|1-e^{-\beta
\sqrt{ (\ell+1)^2R^{-2}+|\Bphi_{pq}|^2}+i\theta_{pq}}\Big|\Big)
\Big\}\,
\label{sbosonic}
.}
Here
\be
|\Bphi_{pq}|=\sqrt{(\Bphi_p-\Bphi_q)^2};\qquad \theta_{pq}=\theta_p-\theta_q.
\ee
The fermion fluctuation determinants contribute
\SP{
S_\text{f}^{(1)} &=-\sum_{pq=1}^{N}
\sum_{\ell=1}^\infty
8\ell(\ell+1)\Big(\frac\beta2\sqrt{
  (\ell+1/2)^2R^{-2}+|\Bphi_{pq}|^2}\\ &\qquad
+\log\Big|1+e^{-\beta
\sqrt{ (\ell+1/2)^2R^{-2}+|\Bphi_{pq}|^2}+i\theta_{pq}}\Big|
\Big)\ .
\label{sfermionic}
} The validity of the one loop potential requires weak 't~Hooft
coupling, $\lambda \ll 1$, but also $\lambda \ll 1/(TR)$. This is
because at $TR \sim \lambda^{-1}$ the spatial $\BS^3$ is the same
size as the nonperturbatively generated magnetic screening scale.
Our objective is now to minimise this potential and analyse how
the ground state depends on the temperature and coupling.
The Casimir terms in \eqref{sbosonic} and
\eqref{sfermionic}\footnote{These are the first terms in the sum over $\ell$,
involving square roots.} can consistently be ignored, since expanding them
in $\Bphi$ gives terms of the form $\beta R|\Bphi|^2(1+{\cal
  O}(R^2\Bphi^2))$ which, at weak coupling,
is always subleading compared with the tree level mass term. It is
important, however, that we do not ignore the $\Bphi_{pq}$
dependence of the exponential terms because these are enhanced at
high temperature $TR \to \infty$, and will play a significant
r\^ole in our analysis in this regime.

\section{Low temperature distribution: $\BS^1 \times \BS^5$}

The action at low $R T \ll 1$ is
\be
S_{TR \ll 1}={N\pi^2 R\beta\over \lambda}\sum_{p=1}^{N}|{\boldsymbol{
\phi}}_p|^2 - \sum_{pq=1}^{N}\log
\left|\sinh \frac{\beta |{\boldsymbol{\phi}}_{pq}|+i
    \theta_{pq}}2\right|-N^2\log2\ .
\ee
This expression
neglects terms in the action \eqref{full} that are exponentially
suppressed at low temperatures. The logarithmic repulsive force
generated between the eigenvalues is a generalised Vandermonde
type interaction resulting from integrating out the off-diagonal
fluctuations; its origin can be traced to the ghost and gauge
fixing terms in the action
\cite{Hollowood:2006xb}. At $\lambda=0$, when the scalars identically
vanish, it reduces to the measure factor in the unitary
matrix model of \cite{Aharony:2003sx}.

The pairwise eigenvalue repulsion is only countered by an external
quadratic attractive potential along the scalar field directions. We
should therefore expect the lowest energy state to be localized in the
scalar directions and maximally spread out, {\it i.e.}~uniformly
distributed, along the $\BS^1$. In this case, using the fact that
\EQ{
\log\Big|\sinh\frac{\beta|\Bphi|+i\theta}2\Big|
=-\log2+\frac{\beta}{2}|\Bphi| -
\sum_{n=1}^\infty\frac1ne^{-n\beta|\Bphi|}\cos(n\theta)\ ,
\label{jhh}
} and noticing that the periodic pieces will average to zero for a
uniform distribution of $\theta_p$s, we have an effective action
\be
S_{TR \ll 1}={N\pi^2 R\beta\over \lambda}\sum_{p=1}^{N}|{\boldsymbol{
\phi}}_p|^2-\frac\beta2\sum_{pq=1}^{N}
|{\boldsymbol{\phi}}_{pq}|\ .
\ee
In fact we will find in the next section that this action is valid
all the way up to the confinement/deconfinement transition. One
way to see this is to note that, as we do in the next section,
that below the confinement/deconfinfement transition the
$\theta_p$s are uniformly distributed around $\BS^1$. Hence, the
averaging over them has the same effect in the action as sending
$T\to0$.

It is convenient to introduce the dimensionless variables
\be
\xb_p ={\beta {\boldsymbol{\phi}}_p} \,.
\ee
At large $N$ we can pass to a continuum limit. The eigenvalues
are described by a joint distribution in seven dimensions,
which satisfies the normalisation condition
\be\label{eq:normalise}
\frac{1}{N} \sum_{p=1}^{N} \to \int d^6 x d\theta\, \rho(\xb,
\theta) = 1 \ ,
\ee
although in the present case $\rho(\xb,\theta)=\rho(\xb)$ only.
The equation of motion for $\xb$ is
\be\label{eq:eom2}
{\pi^2 R T\over \lambda}\xb =\pi\int_{D} d^6 x'\,\rho(\xb')
{\xb-\xb'\over |\xb -\xb'|} \,,
\ee
where the eigenvalue distribution has support in some domain $D
\subset \boldsymbol{R}^6$ and $\xb \in D$.

We can see that \eqref{eq:eom2} has a solution for which $D$ is
an $\BS^5$ of radius $r$, which is easily determined. The density is
found to be
\be
\rho(\xb)= \frac{\delta(|\xb|-r)}{2 \pi^4 r^5} \,,
\ee
where the radius $r$ is
\be\label{eq:lowTs5}
r = \frac{\lambda}{\pi^3 R T} \frac{1024}{945} \,.
\ee

This solution coincides with the zero temperature $\BS^5$ solution of
\cite{Berenstein:2005aa} which is what we should expect in the low
temperature limit where thermal effects are negligible.
Importantly, the smooth $\BS^5$ is a consequence of the large $N$
limit. The $SO(6)$ symmetry of the ${\cal N}=4$ theory is left
unbroken in this saddle by a global $SO(6)$ rotation on each of
the ${\Bx}_p$ combined with the action of the Weyl group of
$SU(N)$ which acts through permutations of the ${\Bx}_p$.

Let us determine the action associated to this $\BS^5 \times
\BS^1$ geometry of eigenvalues
\EQ{
S_{\BS^5\times \BS^1}={2N^2\pi^3TR\over\lambda}\int d^6 x \;
\rho(\xb) \;\xb^2
-4\pi^2 N^2 \int d^6 x \;\;\int d^6x'\;
\rho(\xb)\rho(\xb')\;\log|\xb -\xb'|\ .
}
To this order in perturbation theory, we find that the
action for this configuration is
\be
\frac{1}{N^2} S_{\BS^5\times \BS^1} = -{\lambda
  \over \pi^4}{1024^2\over 945^2} \frac{1}{TR} \,.
\ee
The action is lower than that of both the solution in which the
scalars have zero
expectation value, and also of the band solution found in
\cite{Hartnoll:2006pj}, in which all ${\Bx}_p$ acquired non-zero expectation
values along the same direction in $SO(6)$ space, thus only
preserving an $SO(5)$ subgroup of the global symmetry group. It
seems very likely that this maximally symmetric saddle is the
absolute minimum of the effective potential at low temperatures,
within the sector of the theory we are considering. We prove
stability against R symmetry breaking perturbations in Appendix A.
Furthermore, in Appendix
\ref{sec:saddles} we find various other saddle points with
reduced R symmetry and find that they have a higher action.

It is striking that the eigenvalue distribution localises to a
hypersurface and does not spread out in all the six noncompact
dimensions. In fact,
we can exclude six dimensional solutions to the effective
equation of motion for the scalars (assuming a uniform distribution on
$\BS^1$), by repeated application of the $\boldsymbol{\nabla}$ operator in six
dimensions. In particular, acting on equation \eqref{eq:eom2} with the
operator $\boldsymbol{\nabla}^2(\boldsymbol{\nabla}
\cdot)$, we find $\int d^6 x'
\,\rho({\Bx}')/|{\Bx}-{\Bx}'|^3=0$ which implies a vanishing
density in six dimensions. This then implies that the eigenvalues
at large $N$ must be constrained to lie on a hypersurface in six
dimensions and this is indeed what we find. An argument along
similar lines was shown in \cite{Berenstein:2005aa}, implying that
eigenvalue distributions arising from similar commuting matrix
models were always singular. Localisation to a hypersurface is the
simplest way to achieve this.

Note that our analysis is consistent with \cite{Aharony:2003sx} when
$\lambda=0$: the emergent $\BS^5$ disappears and only the configuration
space of eigenvalues of the Polyakov loop remains. As advocated in
\cite{Berenstein:2004kk, Berenstein:2005aa} the large $N$ vacuum configuration of
eigenvalues should be interpreted as the emergence, at weak 't
Hooft coupling in our case, of the $\BS^5$ in the dual $\boldsymbol{AdS}_5 \times
\BS^5$  geometry. The connection to a dual geometry is discussed in
more detail below.

\section{The deconfinement transition: $\BS^1 \times \BS^5 \to \BS^6$}

This section describes the distribution for $TR\ll
\lambda^{-1/2}$. In particular, we will be interested in the
`deconfinement' or `Hagedorn' transition that occurs at $TR_c
\approx 0.38$ \cite{Aharony:2003sx}. We will see that the
eigenvalue distribution undergoes a topology change: $\BS^1 \times
\BS^5 \to \BS^6$. We will comment on possible spacetime
interpretations of this transition in a later section.
The details of the transition are accessible within
standard perturbation theory in $\lambda$ which is valid for all
temperatures $TR \ll \lambda^{-1/2}$.
%When $TR$ becomes comparable to $\lambda^{-1/2}$,
%the size of the $\BS^3$ approaches the electric or Debye screening
%length $R \sim (\sqrt\lambda T)^{-1}$ and above these temperatures,
%resummed perturbation theory in $\sqrt\lambda$ becomes applicable.

To analyze the vicinity of the phase transition we make use of
certain properties of the equilibrium distribution of the scalar
fields at weak coupling. At an equilibrium, the tree level
attractive potential balances out the pairwise repulsive forces
generated at one loop. It is straightforward to see that the one
loop contribution to the effective action \eqref{full} can only
compete with the tree level term if $|\xb|\sim{\cal O}(\lambda)$.
Hence, in order to be consistent with the perturbative expansion
we should expand the one loop contribution in powers of $\xb$.
This has the effect of reorganizing the pertubative expansion. The
only caveat is that when the temperature is very high, $TR
\gtrsim \lambda^{-1/2}$, a further reorganization occurs which we
shall describe in a later section. In this section we shall assume
that $TR \ll
\lambda^{-1/2}$.

We now define $\xb = \lambda \tilde \xb$ and expand the action up to
${\cal O}(\lambda)$ to obtain
\be
S = \lambda^0 S^{(0)}(\theta)+
\lambda\left( S^{(1)}(\theta,\tilde \xb)+S_\text{2-loop}(\theta)\right)
+{\cal O}(\lambda^2)\ ,
\label{hyy}
\ee
where
\be
S^{(0)}[\theta]= \sum_{pq=1}^{N} \sum_{n=1}^\infty
\frac{1}{n} \Big[1 - z_B(e^{-n/TR}) - (-1)^{n+1} z_F(e^{-n/TR})
 \Big] \cos(n\theta_{pq}) \  . \label{totap}
\ee
This is precisely the $\lambda=0$ effective potential of
\cite{Aharony:2003sx}.  The single particle partition functions $z_B$
and $z_F$ will be
given below. At the Hagedorn transition, the coefficient of the
lowest cosine term becomes negative \cite{Aharony:2003sx}. The one
loop term in the action \eqref{hyy}, using \eqref{jhh}, is given by:
\be
S^{(1)}[\theta,\tilde \xb] = N\pi^2 TR
\sum_{p=1}^{N}|\tilde \xb_p|^2-\frac12
\sum_{pq =1}^{N}
|\tilde
\xb_{pq}|\Big[1+2\sum_{n=1}^\infty\cos(n\theta_{pq})\Big]
\ . \label{totap2}
\ee
It is clear from the form of the expansion \eqref{hyy} that to
leading order the $\theta_p$s are unaffected by the scalars. In
other words the $\theta_p$, and their density in the large $N$
limit, behave to leading order exactly as they did with zero
scalar VEVs. The scalar VEVs themselves are then, to leading
order, determined by finding the minimum of $S^{(1)}(\theta,\tilde
\xb)$ with the given values for $\theta_p$. Notice that although
this term is of the same order as a two loop contribution, the
latter contribution only involves the $\theta_p$, as indicated,
and therefore doesn't contribute to the leading $\tilde {\Bx}$
distribution.

The term in square brackets in \eqref{totap2} is simply the delta
function $\pi\delta(\theta_{pq})$, restricted to even functions.
Note that the $\theta$ eigenvalue density is indeed an even
function \cite{Aharony:2003sx}. Taking the large $N$ limit and
describing the eigenvalues by the joint density
$\rho(\theta,\tilde \xb)$, as previously, we have
\be
\frac{1}{N^2} S^{(1)} = \pi^2 TR \int
 d\theta\,d^6\tilde x\,\rho(\theta,\tilde \xb)
|\tilde \xb|^2- \pi
\int  d\theta\,d^6\tilde x\,
\,d^6\tilde x'\,\rho(\theta,\tilde \xb)\rho(\theta,\tilde \xb')
|\tilde \xb - \tilde \xb'|\ .
\ee
We now proceed to minimise this action.

The fact that the reduced density
\be
\rho(\theta)=\int d^6\tilde x\,\rho(\theta,\tilde \xb)\ ,
\ee
is determined by the zero scalar VEV problem, together with
preserving $SO(6)_R$ symmetry, implies that the joint density is
\be
\rho(\theta,
\tilde \xb)=\frac{\rho(\theta)\delta(|\tilde \xb|- r(\theta))}{
|\tilde \xb|^5 \sqrt{1+ (dr/d\theta)^2}\text{Vol}\,\BS^5}\ ,
\label{jde}
\ee
where the unknown function $r(\theta)$ determines the size of the
$\BS^5$ as it is fibred over the support of $\rho(\theta)$. We
recall that that we have already argued that other $SO(6)$
symmetric configurations for which the scalar field spectral
density has a smooth six dimensional support are not allowed, as
in \cite{Berenstein:2005aa}. From \eqref{jde} we have
\EQ{
\frac{1}{N^2} S^{(1)}= \pi^2 R T \int
 d\theta\,\rho(\theta) r(\theta)^2- \pi\sqrt2 C
\int  d\theta\,\rho(\theta)^2 r(\theta)\ ,
} where
\be
C = \frac{2048 \sqrt{2}}{945\pi} \,.
\ee
Completing the square gives
\be
\frac{1}{N^2} S^{(1)} = \pi^2 R T \int
 d\theta\,\Big[\rho(\theta)
\Big(r(\theta)-\frac{C}{\sqrt2\pi
   TR}\rho(\theta)\Big)^2-\frac{C^2}{2\pi^2(TR)^2}\rho(\theta)^3\Big]\ .
\ee
The final term only depends on $\theta$ and simply contributes to
the two loop order distribution of the $\theta_p$s. Therefore it
can be ignored for our purposes. However, it will potentially
influence higher loop computations for determining the order(s) of
possible phase transitions at finite coupling
\cite{Aharony:2003sx}

Hence, for a minimum we have
\be
\fbox{
$\displaystyle r(\theta)=\frac{C}{\sqrt2\pi R T}\rho(\theta)\ .
\label{gtr}$
}
\ee
This result directly connects the shape of the eigenvalue
distribution and the eigenvalue density, and implies that the
topology change $\BS^5 \times \BS^1\to \BS^6$ occurs as $T$ is increased
through the phase transition. This is because $\rho(\theta)$
changes from the uniform to a gapped distribution with support
$-\theta_0<\theta<\theta_0$, where $\rho(\pm\theta_0)=0$
\cite{Aharony:2003sx}. When $\rho(\theta)$ is uniform, the
distribution is $\BS^1 \times \BS^5$, as in the previous section. Once
$\rho(\theta)$ becomes gapped, then the distribution is an $\BS^5$
fibred over an interval with the size of the $\BS^5$ vanishing at
the endpoints. This is topologically an $\BS^6$.

We can now check the consistency of our expansion in $\xb$ by
investigating whether the higher terms in the expansion are finite
on the solution \eqref{gtr}. The potentially dangerous higher
terms, coming from expanding $e^{- n |\xb_{pq}|}
\cos(n \theta_{pq})$, have the form
\SP{
&\sum_{pq=1}^{N}
\sum_{n=1}^\infty \frac{\lambda^{m+1}}{n}\big| n \tilde \xb_{pq}\big|^{m+1}\cos(
n\theta_{pq})  \\
& \thicksim \int_0^\pi d\xi\sin^4\xi\int_{-\theta_0}^{\theta_0}
d\theta \,\partial_\theta^m\Big[
\rho(\theta)\rho(\theta')(\rho(\theta)^2
+\rho(\theta')^2-2\rho(\theta)\rho(\theta')\cos\xi)^{(m+1)/2}
\Big]_{\theta'=\theta}\ .
\label{pop}
}
Here, we used the fact that
\be
\sum_{n=1}^{\infty} n^m \cos (n\theta) = \pi \partial_\theta^m
\delta(\theta) \,,
\ee
and integrated by parts. The question is whether the terms
\eqref{pop} are finite. The potential problem occurs at the edges
of the distribution $\theta=\pm\theta_0$ where there are possible
singularities. We know \cite{Aharony:2003sx}, see also the
following section, that in the vicinity of the edge
$\theta=\theta_0$ the density behaves as $\rho(\theta)\sim
\sqrt{\theta_0-\theta}$. It is not difficult to show that the
integrals are completely regular at $\theta=\theta_0$. Hence the
analysis is consistent across the transition and into the high
temperature phase (assuming $(TR)^2\ll\ 1/\lambda)$.

In summary, inclusion of the lowest order corrections in $\lambda$
leads to an interpretation of the Hagedorn/deconfinement transition as a
topology changing transition, $\BS^1\times \BS^5\rightarrow \BS^6$, in
eigenvalue space. The possible implications of this geometric
transition for the dual spacetime will be discussed below.

\section{Intermediate temperatures: $\BS^6$ ellipsoid}

We now determine the joint eigenvalue distribution in
the temperature range $1 \ll TR \ll \lambda^{-1/2}$ and provide
further evidence for the appearance of the $\BS^6$ topology above the
Hagedorn/deconfinement temperature. In this
region, the $\theta_p$ have the Wigner semi-circular distribution.
To see this we have to analyse $S^{(0)}$ for $TR \gg 1$. First of all,
recall that the single particle partition functions in \eqref{totap} are
given by
\cite{Aharony:2003sx}
\EQ{
z_B(x)=\frac{2x(3+6x-x^2)}{(1-x)^3}\ ,\qquad
z_F(x)=\frac{16x^{3/2}}{(1-x)^3}\ . }
For $TR \gg 1$ this implies
\EQ{
z_B(x)\to 16(TR)^3/n^3 \ ,\qquad z_F(x)\to16(TR)^3/n^3 \ , }
and we expect that the $\theta_p$ are small at high temperatures,
in which case using $\sum_{n=1}^\infty(2n-1)^{-2}=\pi^2/8$, we
have in this limit
\be
S^{(0)} \to -\sum_{pq=1}^{N}\Big(\log|\theta_{pq}|-2\pi^2(TR)^3
\theta_{pq}^2\Big)
=-\sum_{pq=1}^{N}\log |\theta_{pq}| +
4N\pi^2(TR)^3\sum_{p=1}^{N}\theta_p^2\ .
\ee
This is the action of the conventional Hermitian matrix model with
quadratic potential. The saddle point equation is
\be
\sum_{q (\neq p)=1}^N\frac{1}{\theta_p-\theta_q}=4N\pi^2(TR)^3\theta_p
\label{kii}
\ee
In the large $N$ limit we can solve for the density in the usual
way, by introducing the resolvent
\be
\omega(x)=\frac1N\sum_{p=1}^{N}\frac1{x-\theta_p}\ ,
\label{yxx}
\ee
in terms of which \eqref{kii} becomes
\be
\omega(\theta+i\epsilon)+
\omega(\theta-i\epsilon)=8\pi^2(TR)^3\theta\ ,
\label{ja}
\ee
for $-\theta_0\leq\theta\leq\theta_0$ and $\epsilon$ is an
infinitesimal which imposes on the left hand side a principal
value. It follows from \eqref{yxx} that
\be
\omega(\theta+i\epsilon)-\omega(\theta-i\epsilon)
=-2\pi i\rho(\theta)\ , \label{jb}
\ee
where again $-\theta_0\leq\theta\leq\theta_0$. It then follows
from \eqref{ja} and \eqref{jb} that the resolvent $\omega(x)$ is
an analytic function of $x$ which has a square root branch cut
between $x=\pm\theta_0$. Since $\omega(x)$ must go to zero for
large $|x|$ this determines uniquely
\be
\omega(x)=4\pi^2(TR)^3\Big(x-\sqrt{x^2-\theta_0^2}\Big)\ .
\ee
By taking the discontinuity across the cut we obtain
\be
\rho(\theta)=4\pi(TR)^3\sqrt{\theta_0^2-\theta^2}\ .
\label{dis}
\ee
Finally, normalising the density gives
\EQ{
\theta_0^2=\frac1{2\pi^2(TR)^3}\ .
}

Using \eqref{gtr} we see that in this region of intermediately
high temperatures the combined density has support on an ellipsoid
with topology $\BS^6$ given by
\be\label{eq:interellipse}
\fbox{
$\displaystyle \frac{\pi^2}{\lambda^2 4 C^2 TR} \xb^2 + 2 \pi^2
(TR)^3 \theta^2 = 1 \ .$ }
\ee
Given that $TR \sqrt{\lambda} \ll 1$, this ellipsoid is very
elongated in the $\theta$ direction. The eigenvalue density
\eqref{jde} is largest at the highly pointed tips at $\xb=0$.

\section{High temperature distributions: $\BS^6$ versus $\BS^5$}

This section studies the distribution at high temperatures $1\ll
TR\ll 1/\lambda$, which includes $TR\sim \lambda^{-1/2}$. We will
see that in this regime the VEVs of the scalar fields are no
longer determined by the $\theta_p$, but rather the full coupled
system must be considered. The new phenomenon we will find in this
regime therefore could not be seen in previous analysis that
neglected these scalar VEVs.

It is worth remarking that for $TR \gtrsim \lambda^{-1/2}$,
the $\BS^3$ size exceeds the electric or Debye scale, and
perturbation theory in $\lambda$ is replaced by a perturbation
theory in $\sqrt\lambda$ due to infrared effects. Importantly for
us, our one loop effective potential remains unaltered by these
resummations since the associated momentum integrals at high
temperatures happen to be insensitive to IR effects.

The analysis of the previous section, however, will break down at
high temperatures where $TR \gtrsim \lambda^{-1/2}$ for a
different reason. At these temperatures, the distribution of the
$\xb$ will begin to affect the $\theta$s. This is precisely the
region when the $\BS^6$ begins to look spherical in the
dimensionless variables $\xb$ and $\theta$. However it is
important to realise that the analysis in this section has an
overlap with the analysis of the last section when $1\ll TR\ll
\lambda^{-1/2}$.

To see what goes wrong with the earlier analysis look at the
corrections \eqref{pop}. These are
\be
\sum_{pq=1}^{N}
\sum_{n=1}^\infty \frac{1}{n}\big|n \xb_{pq}\big|^{m+1}\cos(
n\theta_{pq}) = \pi \sum_{pq=1}^{N}
\big|\xb_{pq}\big|^{m+1} \partial_\theta^m \delta(\theta_{pq}) \,.
\ee
We can estimate the behaviour of these terms as a function of
$\lambda$ and $TR$ by using the fact that from
\eqref{eq:interellipse} when $TR \gg 1$ we have $\xb \sim \lambda
(TR)^{1/2}$ and $\theta \sim (TR)^{-3/2}$. Hence, the correction
goes like $(\lambda (TR)^2)^{m+1}$. Clearly these corrections
cannot be ignored in the high temperature regime.

However, we expect that when $TR \gg 1$ then the eigenvalue
distribution will satisfy $\theta_p, |\xb_p| \ll 1$. With these
assumptions, the high temperature action is found to be
\be
S_{TR\gg 1}={N\pi^2 TR\over \lambda}\sum_{p=1}^{N} |\xb_p|^2-
\frac12\sum_{pq=1}^{N}\log
\left( |\xb_{pq}|^2+
    \theta_{pq}^2\right) + \pi^2 R^3 T^3 \sum_{pq=1}^{N} \left( |\xb_{pq}|^2 + 2\theta_{pq}^2
    \right)\,.
\label{hta}\ee
The validity of the assumptions is verified a posteriori from the
solution. The large $N$ equations of motion following from this
action may be written
\bea\label{eq:hiTeom}
P \xb = \int d^6 x' \;d\theta'\;\rho(\xb',\theta')\;{
(\xb-\xb')
\over |\xb -\xb'|^2 + (\theta-\theta')^2} \,, \nonumber
\\
Q \theta = \int d^6 x' \;d\theta'\;\rho(\xb',\theta') \;{\theta
- \theta'
\over |\xb -\xb'|^2 + (\theta-\theta')^2} \,,
\eea
where $P = \pi^2 TR (1/\l + 2 R^2 T^2)$ and $Q = 4 \pi^2 R^3 T^3$.
These equations have more than one interesting solution. Once
again, by application of the seven dimensional operator
$\boldsymbol{\nabla}^2(\boldsymbol{\nabla}\cdot)$ in the
$({\Bx},\theta)$ space, we can argue that non-trivial solutions to
the equations of motion must be hypersurfaces in seven dimensions.
For a given $P$ and $Q$ there are two solutions which preserve the
$SO(6)_R$ symmetry. One is topologically $\BS^6$ and the other is
topologically $\BS^5$. We expect one of these maximally symmetric
solutions to have the lowest action. Before discussing the
solutions, we comment on a scaling property of the solutions with
$P$ and $Q$. In appendix C we present various solutions that do
not preserve the full R symmetry.

\subsection{Scaling of the solutions and action}

The effect of changing the coupling or the temperature of the
theory is incorporated in the values of $P$ and $Q$ in the high
temperature equations of motion \eqref{eq:hiTeom}. Under the
rescaling\be
\tilde P = \mu P \,, \qquad \tilde Q = \mu Q \,,
\ee
the equations of motion and the normalisation condition
\eqref{eq:normalise} imply
that the solution remains the same and scales to
\be
\tilde \xb = \frac{1}{\sqrt{\mu}} \xb \,, \qquad \tilde \theta =  \frac{1}{\sqrt{\mu}} \theta \,.
\ee
The action evaluated on the scaled solution becomes
\be
S[\tilde \xb, \tilde \theta] = S[\xb,\theta] + \frac{1}{2} \log \mu \,.
\ee

We can use this scaling property to fix $Q$ and work with $P/Q$.
When we do numerics shortly, it will be convenient to work with $Q=1$.
Even though this value is not in the high temperature regime, we simply
have to scale the solutions as we have just described. Note that
\be
\frac{P}{Q} = \frac{1}{4} \frac{1}{(TR)^2 \lambda} + \frac{1}{2} \,,
\ee
so that $P/Q \to \infty$ corresponds to $(TR)^2\lambda\ll1$,
whereas $P/Q \to 1/2$ corresponds to high temperatures
$(TR)^2\lambda\gg1$ with the caveat $(TR)\lambda\ll1$.  The
solution in this regime only depends on the combination
$(TR)^2\lambda$ and so, increasing or decreasing this parameter
can also be interpreted as varying the value of the 't~Hooft
coupling $\lambda$, at a fixed temperature and at weak coupling.

\subsection{$\BS^5$ solutions}

These have $\theta = 0$ and preserve the full R symmetry group of the
theory in the large $N$ limit. We can write the eigenvalue density
\be\label{eq:s5density}
\rho(\xb, \theta) = \frac{\d(\theta)\d(|\xb| - r)}{\pi^3 r^5}
\,,
\ee
where the radius is
\be\label{eq:s5radius}
r = \frac{1}{\sqrt{2 P}} \,.
\ee
Evaluating the action on the solution gives
\be\label{eq:s5action}
\frac{1}{N^2} S_{\BS^5} = \frac{5}{24} + \frac{1}{2} \log 2 P \,.
\ee
These solutions are fully collapsed in the $\theta$ direction,
$\rho(\theta)=\delta(\theta)$. At low temperatures, these are unstable
saddle point configurations, but play an important role at high temperatures
as we show below.

\subsection{Ellipsoidal $\BS^6$ solutions}

We now find the solutions which preserve $SO(6)_R$ symmetry and
where the eigenvalues spread out into a (closed) six dimensional
surface in ${\mathbb R}^7$. These are the high temperature
continuations of the ellipsoids we found at intermediate
temperatures.

We can parametrise the surface by
\be
\theta = f(r) \,,
\ee
where as before (although now without the factor or $\lambda$) $r
= |\xb|$. The induced metric on this surface is
\be
ds^2=(1+f'(r)^2) dr^2 + r^2 (d\phi^2 + \sin^2\phi\,d\Omega_4^2).
\ee
Thus, for instance, for a sphere of radius $A$ we have $f(r)=\pm
\sqrt{A^2-r^2}$ while for an ellipsoid with semi-axes $A$ and $B$,
$f(r)=\pm B \sqrt{1- r^2/A^2}$. Note that the parametrisation we
are using forces us to consider the solution in terms of two
separate components. We now proceed to derive an action whose
variation determines $f(r)$.

By $SO(6)$ symmetry, the most general spectral density for the
eigenvalue distribution at large $N$ must be of the form
\be
\rho(\xb, \theta) = g(r)\;\left(\delta(\theta- f(r))+\delta(\theta+f(r)\right)
\ee
where $f(r)$ is taken to be positive. We can now determine the
action restricted to the $SO(6)$ symmetric ansatz. As we will be
reducing the problem to one dimension, it is convenient to
introduce the effective spectral density
\be\label{eq:Gr}
G(r)=2 \pi^3 r^5 \sqrt{1+f'(r)^2}\,g(r) \,.
\ee
The high temperature action is therefore rewritten as
\bea\label{eq:action1}
\frac{1}{N^2} S &=& \int dr\, G(r)(P r^2+ Q f(r)^2)- {2
\over 3 \pi} \int dr\, \int dr' \, \int_0^\pi d\phi \sin^4\phi\,
\\\nonumber\\\nonumber
&\times&  G(r)\, G(r') \log\left({|{\bf r} - {\bf
r}'|^2+(f(r)-f(r'))^2 }\right)\left({|{\bf r} - {\bf
r}'|^2+(f(r)+f(r'))^2 }\right)\,.
\eea
We should also add a Lagrange multiplier term that implements the
normalisation constraint: $\mu \left(\int dr G(r) - 1
\right)$. By functionally differentiating with respect to $G$ and
$f$ we can show that this action consistently leads to the
equations of motion restricted to our $SO(6)$ ansatz.

It is possible to perform the $\phi$ integral in
\eqref{eq:action1} using contour integration. The answer is
\bea\label{eq:1daction}
\frac{1}{N^2} S & = & \int dr\, G(r)(P r^2+ Q f(r)^2)-
\frac{1}{2} \int dr\, \int dr' \, G(r)\, G(r') \left[\log rr' + \frac{7}{12} - \log 2 \right. \nonumber \\
& + &  \left. \log \left(\sqrt{K^2+1} + \sqrt{K^2-1} \right)
- \frac{2}{3} \frac{3 K^2 \sqrt{K^4-1} + 4 (K^4-1)}{(\sqrt{K^2+1}+\sqrt{K^2-1})^4} + f(r') \to - f(r')
\right] \nonumber \\
& + & \mu \left(\int dr G(r) - 1 \right) \,,
\eea
where
\be
K^2 = \frac{r^2 + r'^2 + (f(r) - f(r'))^2}{2 r r'} \,.
\ee
In \eqref{eq:1daction} we have reduced the problem to one
dimension. Variation of this action leads to a pair of integral
equations for $f(r)$ and $G(r)$. We have solved these equations
numerically using a simple Monte Carlo algorithm. The algorithm
discretises the $r$ axis into $N$ points: $\int dr G(r) \to
\sum_{i=1}^N$ and then minimises the multiparticle action
\eqref{eq:1daction} by relaxation. A test of our code is that at
the point with enhanced $SO(7)$ symmetry, when $P=Q$, it correctly
reproduces a six sphere of radius $1/\sqrt{2 P}$. The action of
this configuration may be found analytically to be
\be
\left. \frac{1}{N^2} S_{\BS^6} \right|_{P=Q} = \frac{107}{120}
+ \frac{1}{2} \log \frac{P}{2}  \,.
\ee
The numerically obtained action agrees with this exact result to
three significant figures, even with a  fairly small number of
points, $N \sim 60$ or so. Our main interest is for $P \neq Q$,
for which we currently do not have analytic results.

The equations that we are solving numerically are very similar to
those considered in the recent paper \cite{Berenstein:2007wz},
which was in a supersymmetric context. The two main differences
are firstly that we are looking for the semiclassical, $N \to
\infty$, saddle point, whereas \cite{Berenstein:2007wz} perform a
quantum Monte Carlo simulation to study the finite $N$
wavefunction. Secondly, we have restricted to an $SO(6)$ invariant
ansatz before applying numerics. This gives a more efficient use
of eigenvalues, which are diluted in fewer dimensions. However, it
means that the $SO(6)$ symmetry is a (consistent) input
assumption. The numerics in \cite{Berenstein:2007wz} did not
assume this symmetry but found it in their results. This provides
the complementary information that the $SO(6)$ ansatz is stable
against $R$ symmetry breaking perturbations.

Figure 2 shows the eigenvalue distributions obtained for three
values of $P/Q$. The form is roughly what we could have
anticipated, given that as $P$ is increased the external force
pushing the eigenvalues to small $r$ becomes greater. The
eigenvalue distributions appear to be ellipsoids to a very high
degree of accuracy. This is perhaps not surprising given our
results from the intermediate temperature regime. The numerics are
not accurate enough to reliably read off a density profile.

The gap in the data points at small $r$ in figure 2 is due to the
fact that the effective density, $G(r)$, is very low in this
region. This is due to the $r^5$ term in \eqref{eq:Gr}. Note
however that there is a data point at $r=0$ for all three curves.
The solid lines in figure 2 are the ellipses determined by the
extent of the distribution in the $r$ and $f$ directions. These
curves match all the numerically obtained data points to two
significant figures.

\begin{figure}[h]
\begin{center}
\epsfig{file=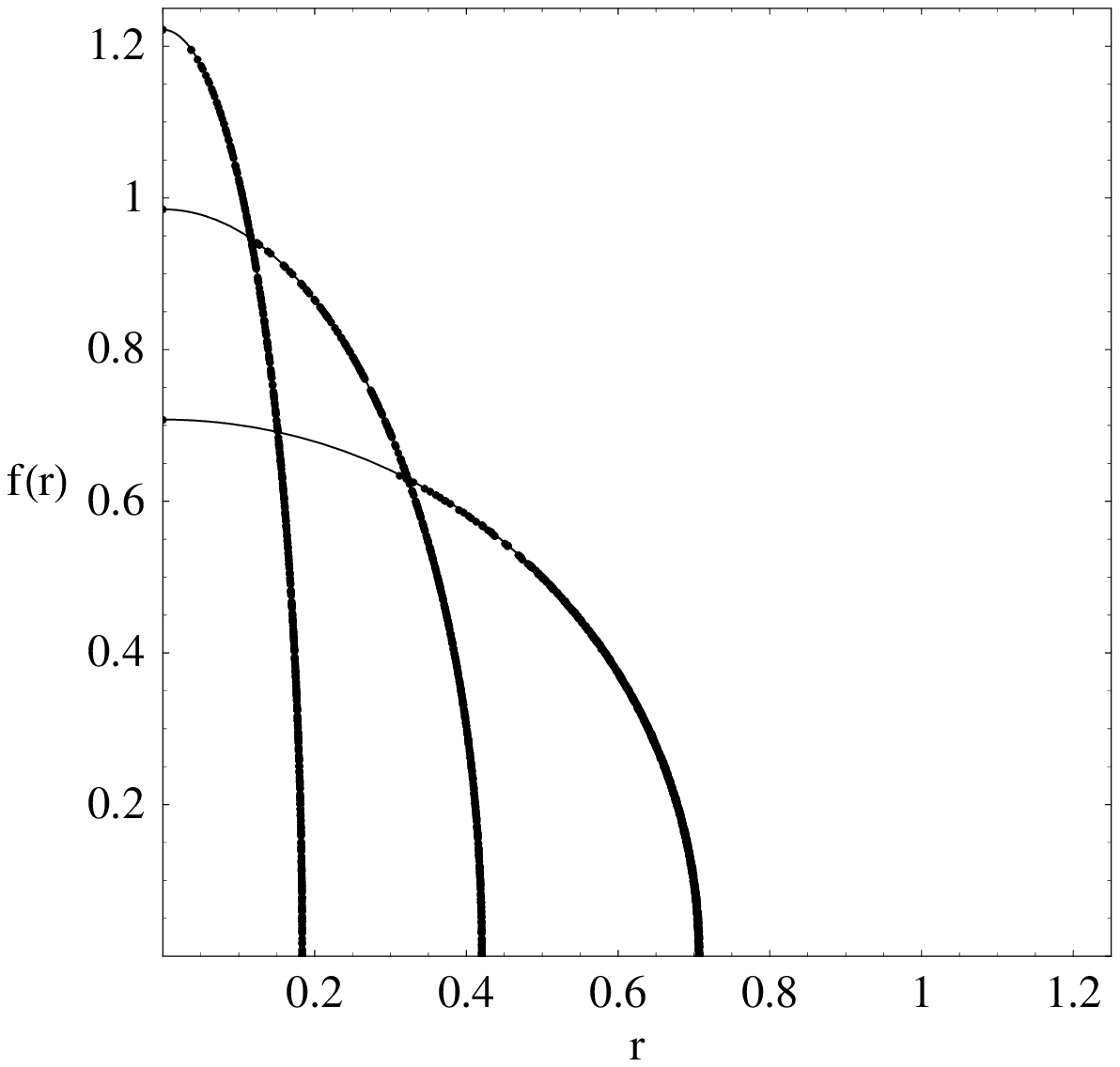,width=7cm}
\end{center}
\noindent \small {\bf Figure 2:} Output from the Monte Carlo
simulation for $\{P,Q\}= \{5,1\}, \{2,1\},\{1,1\}$ with
$N=500,500,1000$ points, respectively. The solid lines are
ellipses with axis ratio determined by the data.
\label{threelines}
\end{figure}

Given the appearance of ellipsoids, a rescaling-invariant question
we can ask is how the axis ratio of the ellipsoid varies with
$P/Q$. Figure 3 is a plot showing the axis ratio at several values
of $P/Q$. There appears to be a linear relation between these
quantities to good approximation over the range plotted. Figure 3
also shows a linear fit to the data points for $P/Q \geq 2$. The
best fit line is
\be\label{eq:bestfit}
\frac{a}{b} = -0.81 + 1.51 \frac{P}{Q} \,,
\ee
where $a$ and $b$ are the lengths of the axis in the $f$ and $r$
directions, respectively.

\begin{figure}[h]
\begin{center}
\epsfig{file=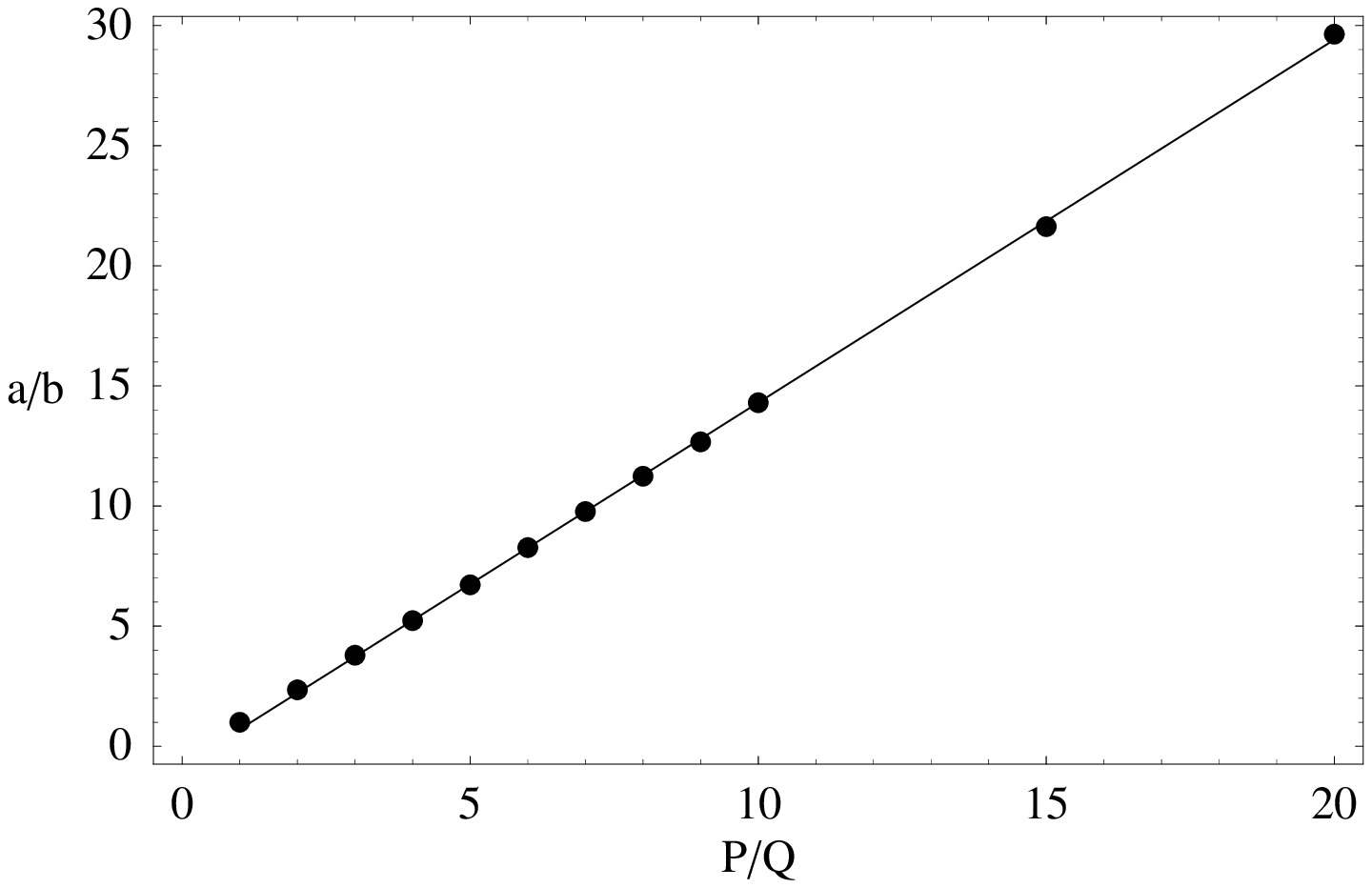,width=10cm}
\label{linefit.eps}
\end{center}

\noindent \small {\bf Figure 3:} The axis ratio of the ellipsoidal
distribution as a function of $P/Q$, with $Q=1$. The data points are
from the Monte Carlo numerics whereas the solid line is a linear
fit.

\end{figure}

In the limit where $\lambda (TR)^2 \ll 1$, or $P/Q$ large, we can
compare the axis ratio with that of the ellipsoid we found in the
intermediate temperature regime \eqref{eq:interellipse}. In terms
of $P/Q$ it follows from \eqref{eq:interellipse} that the axis
ratio for those ellipsoids is also linear in $P/Q$
\be
\frac{a}{b} = \frac{945 \pi}{2048} \frac{P}{Q} \approx 1.45
\frac{P}{Q} \,.
\ee
This is fairly close to \eqref{eq:bestfit}, especially given that
the fit \eqref{eq:bestfit} included a range of points outside the
region of validity of \eqref{eq:interellipse}. A more precise
matching comes from taking, for instance, the value $P/Q = 20$.
This corresponds to $\lambda (TR)^2 \sim 1/80$, and so should be
well described by the intermediate temperature analysis. For this
value we find that the numerical axis ratio is 29.6 whereas the
analytic result is 29.0. This close agreement is a test of both
the numerics and analytic results.

Finally, we can compute the action of the solution as a function
of $P/Q$. The result is shown in figure 4 for $Q=1$. We have also
included in the plot the action of the round $\BS^5$ solution and
the action of the two dimensional ellipse solution discussed in
\cite{Hartnoll:2006pj}.

\begin{figure}[h]
\begin{center}
\epsfig{file=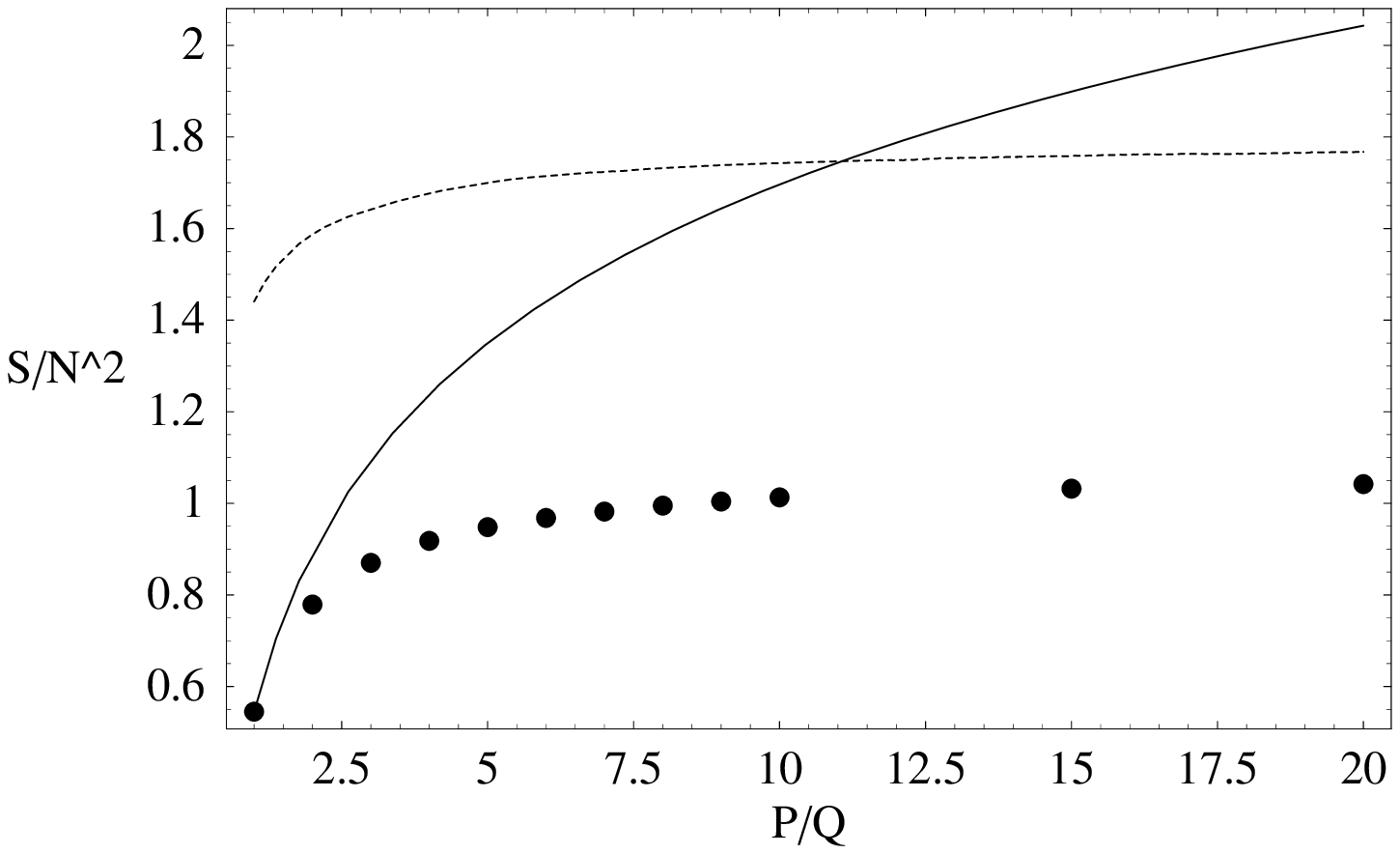,width=10cm}
\end{center}

\noindent \small {\bf Figure 4:} The action of three different eigenvalue
configurations as a function of $P/Q$ with $Q=1$. The data points
are from the $\BS^6$ ellipsoid. The solid curve is the round $\BS^5$
solution. The dotted curve is the two dimensional ellipse
solution.

\end{figure}

As with the axis ratio, at large $P/Q$ we can compare the
numerically obtained actions with the analytic result in the
intermediate temperature regime. At $P/Q=20$ the numerically found
action is 1.04 and the action of the ellipsoid
\eqref{eq:interellipse} is 1.09.

We see that throughout the range plotted, $P/Q \geq 1$, the
ellipsoidal eigenvalue distribution has the lowest action and
therefore dominates the partition function. Therefore, at
sufficiently weak coupling, the ellipsoidal distributions describe
the vacuum of the theory. However, as $P/Q$ is lowered past $1$,
recall that lowering $P/Q$ corresponds to either increasing the
temperature at fixed weak coupling or increasing the (weak)
coupling at fixed temperature, it looks like the curves for
$\BS^5$ and the six dimensional ellipsoid might cross. This raises
the prospect of a new phase transition in the high temperature
regime as a function of coupling.

\section{A second order phase transition: $\BS^6 \to \BS^5$}

In figure 4 we saw that as the coupling is increased at fixed high
temperature, the action for the $\BS^5$ solution comes closer and
closer to that of the, increasingly anisotropic,  ellipsoidal
$\BS^6$ solution. At the point with enhanced $SO(7)$ where we have
analytic results, $P=Q=1$, the difference between the actions is
surprisingly small, given that $P/Q=1$ is not particularly close
to the asymptotic regime $P/Q \to 1/2$
\be
\frac{1}{N^2} \left[ S_{\BS^6} - S_{\BS^5} \right]_{P=Q=1} \approx - 0.0098 \,.
\ee
What happens as $P/Q$ is decreased below this value? We now give
four arguments that, taken together, strongly suggest that a
second order phase transition occurs in the commuting saddle at
$\lambda (TR)^2 = 1$, or $P/Q = 3/4$. This is the curve that we
included in figure 1 above.

\subsection{The actions converge}

Figure 5 shows the action from the numerical computation together
with the analytically computed action of the round $\BS^5$
solution \eqref{eq:s5action} as a function of $P/Q$, for $0.7 \leq
P/Q \leq 1.2$. We see that to within the accuracy of the numerical
computation, the error seems to be under $0.5\%$, the numerical
data joins the curve of the $\BS^5$ action somewhere around $P/Q
\sim 0.75$ or $0.8$ and then follows this curve for lower $P/Q$.

\begin{figure}[h]
\begin{center}
\epsfig{file=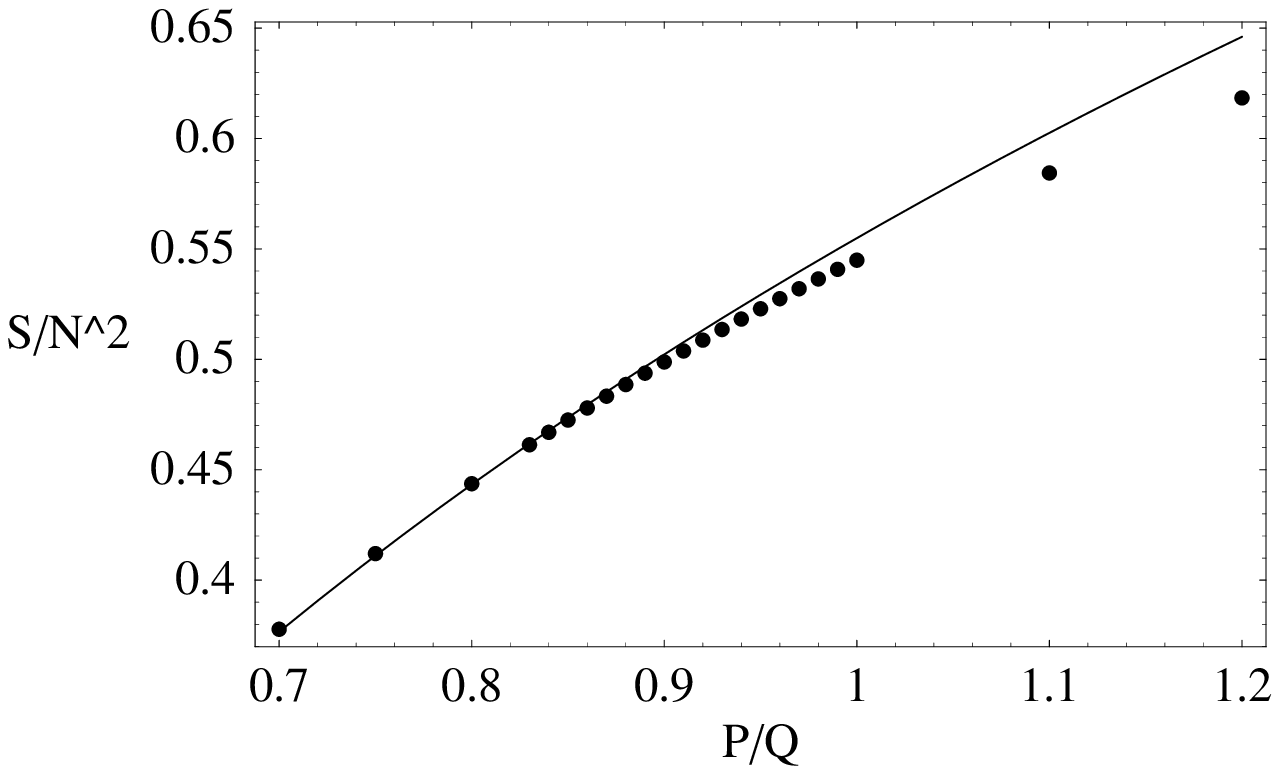,width=10cm}
\end{center}

\noindent \small {\bf Figure 5:} The data points are the numerically computed action (with $N=150)$.
The curve is the action of the $\BS^5$ solution.

\end{figure}

The simplest interpretation of figure 5 is that beyond some
critical coupling, such that $P/Q \sim 0.75$ or $0.8$, the $\BS^5$
saddle becomes the lowest action configuration. The Monte Carlo
simulation therefore converges on the $\BS^5$ saddle below this
value. If this indeed occurs, it indicates that there is a quantum
phase transition at the critical coupling in which the vacuum
manifold of eigenvalues changes topology: $\BS^6 \to \BS^5$. It is
a quantum phase transition in the sense that it is not associated
with any symmetry breaking and the different phases have
topological characterisations. It is possible however to write
down non-topological ``order parameters'' that go from non-zero to
zero as we cross the transition, such as $\langle\xb^2\rangle^2 -
\langle\xb^4\rangle$. It should be clear from this order parameter
that the transition will not be visible if the scalars are
integrated out.

The convergence of actions does not tell us whether the possible
transition is first or second order. We will now show analytically
that at $P/Q=0.75$ the $\BS^6$ saddle becomes completely localized
in the $\theta$ direction and becomes a round $\BS^5$.
Furthermore, we we will find zero modes about the $\BS^5$
solution at precisely $P/Q=\tfrac34$
which provide compelling evidence that there is indeed a
second order phase transition.

\subsection{The $\BS^6$ collapses to $\BS^5$}

Motivated by the numerical results above, we now look for an
analytical condition on the parameters $P$ and $Q$, for the $\BS^6$
solution to collapse to an $\BS^5$. The central fact we will use is the
existence of the $\BS^6$ saddle for an appropriate parameter range, a
fact which we have established both numerically and analytically for
a wide range of temperatures. In the temperature range $TR\gg1$, the
$\BS^6$ topology solves the equations \eqref{eq:hiTeom}.

Since the $\BS^6$ solution is actually a round $\BS^5$ fibred over
a finite interval in the $\theta$ direction, we may differentiate
the second equation of motion \eqref{eq:hiTeom} with respect to
$\theta$ to obtain a condition which must be fulfilled by the same
solution\footnote{We may do this consistently for every value of
$\theta$ lying inside  the distribution, and by continuity we can
aslo apply the condition to all points approaching the edges of
the distribution in $\theta$-space.}:
\be\label{eq:diff2}
Q = \int d^6x'\;d\theta'\;\rho({\Bx}',\theta')
{{|{\Bx}-{\Bx}'|^2-(\theta-\theta')^2}\over
\left(|{\Bx}-{\Bx}'|^2+(\theta-\theta')^2\right)^2} \,.
\ee
We emphasize that this will be automatically satisfied by
the $SO(6)$ symmetric $\BS^6$ topology, since the equation of motion
holds at each point on the support of $\rho({\Bx},\theta)$ in the
$\theta$ direction.

We now want to find at what temperature the extent of the
effective distribution in $\theta$ shrinks to zero size, and thus
the $\BS^6$ collapses to an $\BS^5$. In such a limit, we expect
the density function $\rho({\Bx}, \theta)$ to smoothly approach
\eqref{eq:s5density} with the $\BS^5$ radius equal to
$1/\sqrt{2P}$. Substituting these into \eqref{eq:diff2}, we find
\be
{Q\over P }= {4\over 3}.
\ee
It is easy to check that the integral in \eqref{eq:diff2} is well
behaved in the limit that $\theta,\theta'$ approach zero, and that
the limiting value of $\tfrac43$ is approached from below. Hence
the $\BS^6$ topology ceases to exist for $Q/P >\tfrac43$ which
translates to $TR>\lambda^{-1/2}$, and at these temperatures with
a fixed weak coupling, the equations of motion are only solved by
the $\BS^5$ configuration of eigenvalues.

We remark that this transition from $\BS^6$ to $\BS^5$ cannot be seen
in the $\lambda=0$ theory, since the critical temperature is
$T=1/({\sqrt\lambda}R)$ , and is driven by the presence of the scalar
expectation values in the $\N =4$ theory.

\subsection{The $\BS^5$ saddle develops a zero mode}

Now consider the stability of the $\BS^5$ solution
\eqref{eq:s5radius}. As mentioned above, we expect the $\BS^5$
solution will be stable against perturbations in the $\xb_p$
directions that would break R symmetry. It is important to check this
and we do so in Appendix B. However, something
interesting occurs if we consider perturbations in $\theta_p$.

Returning for the moment to the discrete system, the quadratic action for fluctuations $\delta \theta_p$ is given by
\be
\delta^{(2)} S = Q N \sum_{p=1}^{N} \d \theta_p^2
- \frac{1}{2} \sum_{pq = 1}^{N} \frac{(\d \theta_p - \delta
\theta_q)^2}{|\xb_{pq}|^2} \,.
\ee
It is straightforwardly shown that in the large $N$ limit, the $N$
eigenvalues are all equal to $N(Q - \tfrac43 P)$. Hence,
the solution is only stable when $Q/P > \tfrac43$, that is
\be
TR>\frac{1}{\sqrt{\lambda}}\ .
\ee
When $Q/P<\tfrac43$ it is unstable to spreading out in the
$\theta$ direction. We have already found the high temperature
configuration with $\theta$ spread out: it is the $\BS^6$ ellipsoid.

The appearance of zero modes in the $\BS^5$ vacuum and negative
modes for $TR<\lambda^{-1/2}$, in conjunction with our earlier
observation that the $\BS^6$ solution merges with the $\BS^5$ at
$TR=\lambda^{-1/2}$, provides clear and solid evidence for a second order
phase transition at this temperature.

We remark that although there are $N$ zero modes at the transition
there is only one $\BS^6$ vacuum due to the action of the Weyl group
permutations on the joint eigenvalues $(\Bphi_p,\theta_p)$.

\subsection{The eigenvalue distributions converge}

In further support of this picture, let us look in more detail at
the output of the numerics as $P/Q \to \tfrac34$. The output no longer
resembles the ellipses shown in figure 1 but rather gives a
cluster of points around a radius that is very close to the radius
of the $\BS^5$, $1/\sqrt{2P}$ as found in \eqref{eq:s5radius} above,
and $f=0$. This is what we would expect to see if the density of
eigenvalues of the $\BS^6$ were accumulating on the equator as the
ellipsoid becomes increasingly squashed. In the limit of this
process, the solution becomes an $\BS^5$.

The three graphs in figure 6 illustrate this process. As $P/Q \to
\tfrac34$, the
eigenvalues start to cluster around $0.81 \approx 1/\sqrt{2 \times 0.75}$.
As in the previous section, the gaps in the distribution are partly due to
a low effective eigenvalue density \eqref{eq:Gr} as $r \to 0$. However, we
clearly see the gap growing as the eigenvalues become more densely
clustered around the equator of the squashed ellipsoid.

\begin{figure}[h]
\begin{center}
\epsfig{file=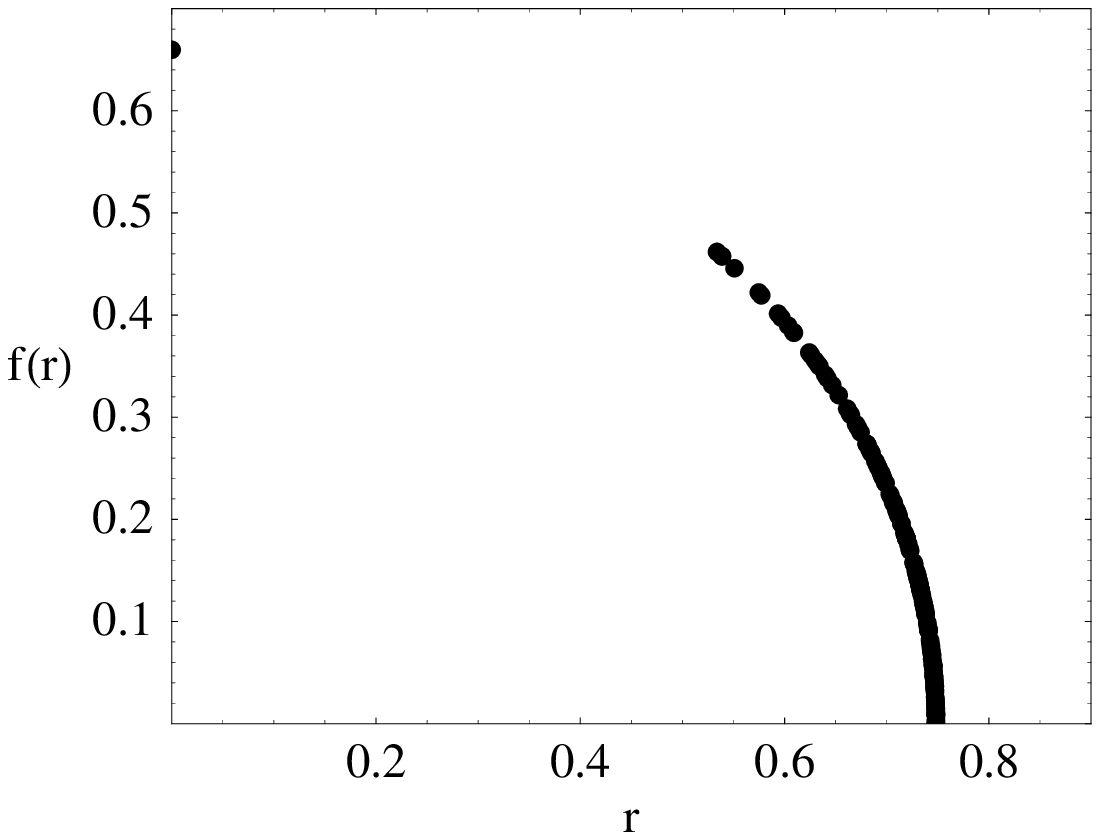,width=4.5cm} \epsfig{file=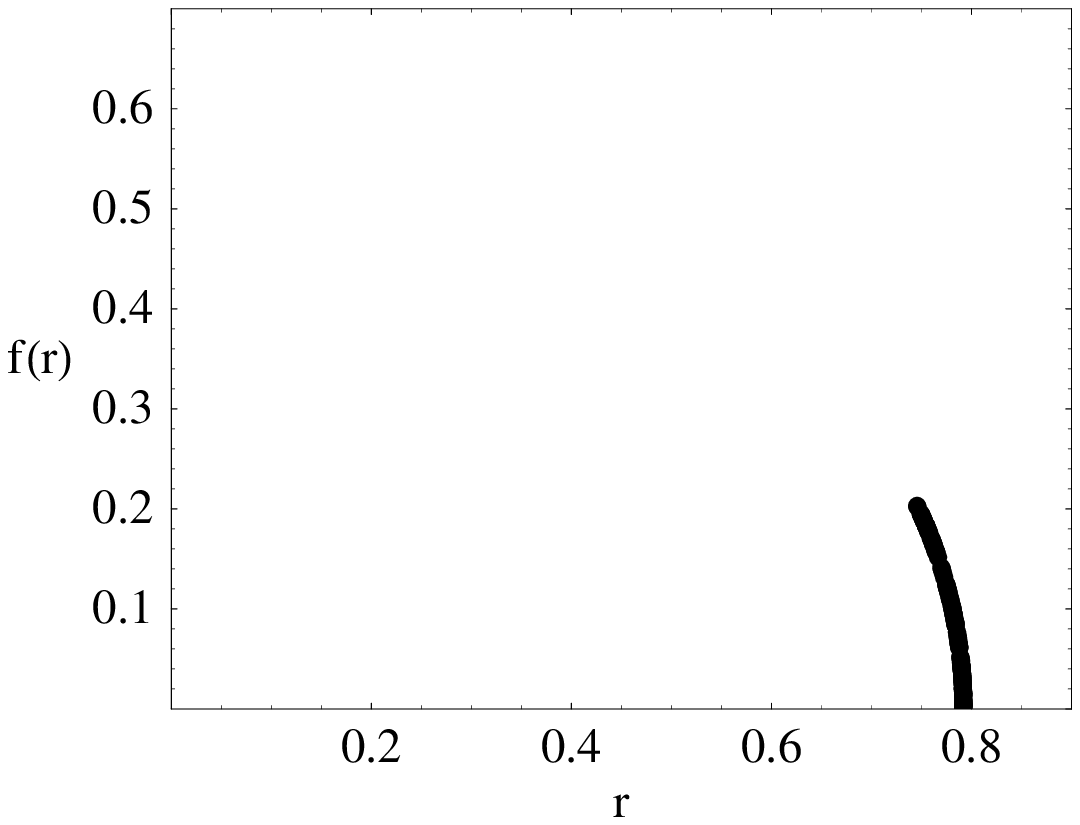,width=4.5cm}
\epsfig{file=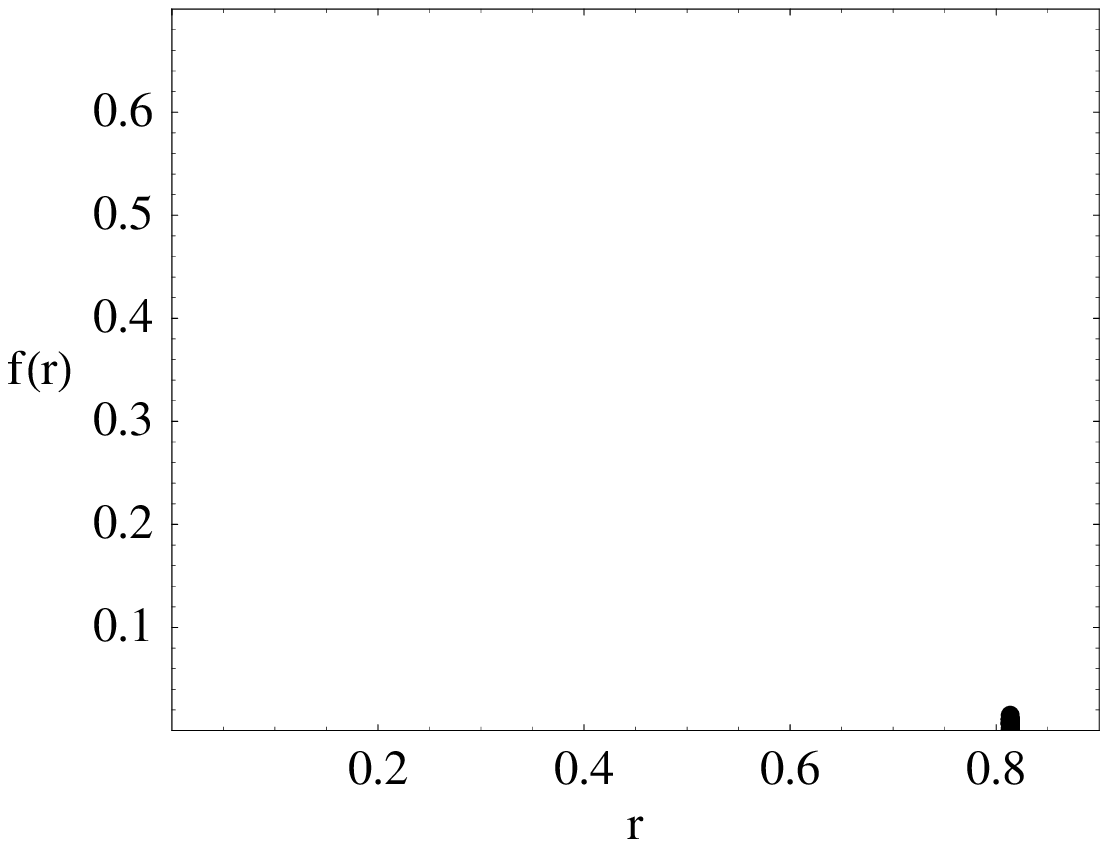,width=4.5cm}
\end{center}

\noindent \small {\bf Figure 6:} From left to right, the
eigenvalue distributions with $P/Q = 0.9, 0.8$ and $0.75$.
Numerically computed with $N = 150$ eigenvalues and $Q=1$.
\end{figure}

\section{Dual spacetime interpretation}

A striking recent development in the supersymmetric sector of the
zero temperature AdS/CFT correspondence is that the scalar field
eigenvalue distribution can be directly connected to the dual
spacetime geometry \cite{Berenstein:2004kk, Lin:2004nb,
Berenstein:2005aa, Berenstein:2007wz}. The most immediate
manifestation of this connection is that the $\BS^5$ part of the
eigenvalue distribution is to be identified with the $\BS^5$ part of
the dual $\boldsymbol{AdS}_5 \times
\BS^5$ geometry \cite{Berenstein:2005aa}. One
can think of the eigenvalue distribution as signalling the fact
that the spacetime geometry has undergone a geometric transition,
in which a noncontractible $\BS^5$ appears, due to the gravitational
backreaction of the $N$ D3 branes. More generally, the eigenvalue
distribution is to be identified with the locus in the dual
spacetime where the $\BS^3$ of the conformal boundary geometry
degenerates
\cite{Berenstein:2005aa, Berenstein:2007wz}. In $\boldsymbol{AdS}_5$ this is
simply the origin of the AdS space.

The question is then whether, in our finite temperature, non
supersymmetric and weak coupling setup, we can reinterpret the
various topology changes undergone by our eigenvalue distributions
as topology changes in a dual spacetime, analogous to the
Hawking-Page transition. Although our saddles are not generically
the dominant saddles, the assumption of commuting matrices does
mean that they are the most `geometric' saddles at weak coupling.
At low temperatures, our $\BS^1 \times
\BS^5$ distribution looks promising. This is indeed the topology
of Euclidean thermal AdS at the origin of the AdS space.

The $\BS^1$ appearing in the low temperature $\BS^1\times \BS^5$
eigenvalue distribution should in fact be associated with the
T-dual of the thermal circle in spacetime, insofar as T-dualising
along a thermal circle is well defined. This is because
eigenvalues of the Wilson line around a circle, in our case the
Polyakov-Wilson loop around the thermal circle, yield the
locations of D branes in the T-dual transverse circle.
%This is because the $A_0$ eigenvalues are associated with the
%Polyakov loop, which describes winding modes about the thermal
%circle.
Supporting this observation, the radius of the $\BS^1$
circle is proportional to the temperature $T$, if we renormalize the
low temperature $\BS^5$ radius \eqref{eq:lowTs5} to unity.

The $\BS^6$ eigenvalue distribution appearing above the first
order Hagedorn transition has a natural explanation if one takes
the notion of T-duality along the thermal circle seriously. In
this picture, the Polyakov loop eigenvalues represent positions of
D2 branes, T-dual to the original D3 branes, along the T-dual
circle. The gapped distribution of the Polyakov loop eigenvalues
above the Hagedorn transition is then a localized distribution of
D2 branes on this circle. A similar localization of D2 branes on a
{\em spatial} circle, at finite temperature, has been argued
\cite{Barbon:2004dd} to produce a near horizon geometry containing a non-contactible
$\BS^6$. In that case, somewhat tantalizingly, an $\BS^1\times
\BS^5 \rightarrow \BS^6$ topological transition of the
Gregory-Laflamme type was predicted from supergravity, and the
localized D2 brane configuration was reinterpreted as winding mode
condensation on the original circle. The interpretation of a
nonuniform eigenvalue distribution as smeared branes has also been
used previously to good effect in studying a field theory dual of
the Gragory-Laflamme black string instability in
\cite{Aharony:2004ig}.

From the weakly coupled field theory perspective, the connection
to a D2 brane setup above the Hagedorn/deconfinement temperature
and below $T=1/(\sqrt\lambda R)$, perhaps suggests that the field theory
is in an effectively three dimensional phase in that region of the
phase diagram. The potential appearance of an $\BS^6$ topology in
the dual spacetime would be consistent with this line of thinking: At
any fixed temperature,
as the 't Hooft coupling is decreased so that string corrections
become substantial, at a critical coupling, when the size of the thermal
circle in string units becomes sufficiently small, the theory undergoes a
stringy transition resulting in the $\BS^6$ topology.

More interesting is the possible dual spacetime interpretation
of the $\BS^5$ topology above the second order transition at
$TR=\lambda^{-1/2}$. It is tempting to identify this topology as the
deep interior of the big black hole in $\boldsymbol{AdS}_5\times \BS^5$ wherein the
thermal circle has shrunk to zero size at the horizon. The simplest phase
diagram implied by the new phase transition, shown in figure 1, also
suggests that this phase might be the continuation to weak coupling of
the big black hole in AdS space.

It is natural to try to associate the disappearance of the product
$\BS^1$ factor in the eigenvalue distribution with the appearance
of a black hole in the dual geometry. However, it is difficult to
make this connection precise. For a start, in black hole
geometries the spatial $\BS^3$ never collapses. The Euclidean
thermal circle degenerates at the horizon instead. The zero
temperature connection of the eigenvalue distribution to the locus
of vanishing $\BS^3$s in the dual spacetime will need to be
modified in the non-supersymmetric thermal situation for a
spacetime interpretation of the eigenvalue distributions to be
possible. Since these distributions characterize the ground state
or the deep IR of the thermal field theory on $\BS^3$, it is
possible that they provide information about the deep interior of
the dual geometry.

\section{Discussion and conclusions}

In this paper we have studied weakly coupled finite temperature
${\mathcal N} = 4$ super Yang-Mills theory on a spatial $\BS^3$,
in the large $N$ limit. The main difference between our work and
previous treatments, \cite{Aharony:2003sx} being the most
important, is that we have considered the effects of condensed
eigenvalues for the six scalar fields of the theory, as well as
the time component of the gauge field, $A_0$. These scalar
eigenvalues have been studied recently in the zero temperature
theory \cite{Berenstein:2005aa, Berenstein:2007wz} where their
condensate is directly related to the dual spacetime geometry. The
scalar eigenvalues condense despite having conformal and thermal
masses because there is a logarithmic repulsion between the
eigenvalues, which at large $N$ overcomes the mass squared terms.

We have only studied certain special saddle points of the full
large $N$ theory, in which the homogeneous modes of the fields $\{
A_0, \Phi_J\}$ condense and commute with each other. Although
these are not generically the dominant saddles, and thus do not
determine the phase structure of the theory, they have several
interesting features. Firstly, they are tractable saddles that
preserve the full $SO(6)_R$ symmetry of the theory. We have seen
that they undergo nontrivial dynamics as a function of temperature
and coupling. More speculatively, because these are the most
geometric saddles at weak coupling, one can wonder whether they
have any connection with the geometry that arises in the strongly
coupled theory via the AdS/CFT duality. For instance, if a
mechanism similar to that described in \cite{Berenstein:2005aa} is
responsible for the emergence of spacetime geometry at finite
temperature, then the strongly coupled theory will be described by
a commuting saddle point. In this scenario, our weak coupling
saddles would be continuously connected to the strong coupling
supergravity geometry.

At low temperatures and through the Hagedorn phase transition into
the deconfined phase, the picture of \cite{Aharony:2003sx} is not
fundamentally modified. The eigenvalue distribution of $A_0$ which
that paper considered becomes part of a higher dimensional
distribution: $\BS^5 \times \BS^1$ at low temperatures and an
ellipsoidal $\BS^6$ above the transition. However, at sufficiently
high temperature $TR=\lambda^{-1/2}$, or alternatively, when the
coupling becomes sufficiently strong (but still small) we found
evidence for a new second order phase transition, in the commuting
saddle. This transition is directly due to the backreaction of the
scalar field eigenvalues on the eigenvalues of $A_0$. Therefore,
this transition could not have been seen without inclusion of the
scalar field eigenvalues. In terms of distributions, at the
transition: $\BS^6 \to \BS^5$.

We presented four pieces of evidence for a second order phase
transition in the weakly coupled theory at $\lambda = 1/(TR)^2$,
with $TR \gg 1$. Firstly, we found numerically that the actions of
the $\BS^5$ and $\BS^6$ saddles meet at this coupling. Secondly,
we showed analytically that the $\BS^6$ solution collapses to
$\BS^5$. Thirdly, the $\BS^5$ solution develops a zero mode at
precisely this point. Finally, we numerically showed how the
$\BS^6$ distribution becomes increasingly squashed and approaches
an $\BS^5$ at this coupling.

An application of the results in this paper is that they will
allow the computation of the Maldacena-Polyakov loop at weak
coupling in the commuting saddles. This is a thermal Wilson loop
involving both the gauge potential and the scalar fields, which
arises naturally in the ${\mathcal{N}}=4$ theory. These loops may
be computed at strong coupling \cite{Hartnoll:2006hr} and are
therefore an interesting observable to compare at weak and strong
couplings. It would be interesting to understand how the $\BS^6
\to \BS^5$ phase transition we have discussed is reflected in the
value of these loops.

\section*{Acknowledgements}

We are especially grateful to David Berenstein for emphasizing to
us the importance of considering all of the six scalar fields. We
would like to thank the anonymous referee and Ofer Aharony for
important critiques of the preprint version of this paper. We are
also happy to acknowledge helpful conversations with Jan de Boer,
David Gross, Gary Horowitz, Carlos Hoyos, Nori Iizuka, Hong Liu,
David Morrison, Asad Naqvi, Joe Polchinski and Steve Shenker. This
research was supported in part by the National Science Foundation
under Grant No. PHY05-51164. U.G. is supported by the European
Commision Marie Curie Fellowship, under the contract
MEIF-CT-2006-039962.

\appendix

\section{Stability of the solutions}

In this appendix, we will consider the stability of some of our
solutions. In particular, we shall consider both the high and low
temperature $\BS^5$ solutions. We only consider the stability to
perturbations in $\Bx_p$ and $\theta_p$, and not to other modes
that we have integrated out.

{\it The high temperature $\BS^5$}

We begin with the high temperature $\BS^5$ solution of Section
6.2. By choosing $\sum_p\Bx_p=\sum_p\theta_p=0$, and scaling the
$\Bx_p$ and $\theta_p$ appropriately, the discrete action
\eqref{hta} has the form
\be
S= -\tfrac12\sum_{pq=1}^N\log\big(|\Bx_{pq}|^2+\theta_{pq}^2\big)
+N\sum_{p=1}^N\big(|\Bx_p|^2+\sigma\theta_p^2\big)
\label{jui}
\ee
with $\sigma=Q/P$, The $\BS^5$ solution corresponds to
$\theta_p=0$ and $|\Bx_p|=1/\sqrt2$. Expanding around the solution
to second order in the fluctuations, we have
$\Bx_p=\tfrac1{\sqrt2}\BOmega_p+\delta\Bx_p$, where $\BOmega_p$ is
a unit 6-vector, we have
\EQ{
\delta S=N\sum_{p=1}^N\big(|\delta\Bx_p|^2+\sigma\theta_p^2\big)
-\tfrac12\sum_{pq=1}^N\Big(\frac{|\delta\Bx_{pq}|^2+\theta_{pq}^2}
{1-\BOmega_p\cdot\BOmega_q}-
\Big(\frac{(\BOmega_p-\BOmega_q)\cdot\delta\Bx_{pq}}
{1-\BOmega_p\cdot\BOmega_q}\Big)^2\Big)+\cdots\ .
\label{frr}
} At large $N$, let us estimate the orders of $N$ of the
off-diagonal relative to the diagonal terms in the quadratic form
of \eqref{frr}. Firstly the off-diagonal terms. One might think
that when $\delta\BOmega=
\BOmega_p-\BOmega_q$ is small these can be large relative the diagonal
terms. The question is: how small can $\delta\BOmega$ be for $N$
points distributed uniformly on $\BS^5$? If it were an $\BS^1$
then this would be $\sim1/N$ and so the sum over pairs in
\eqref{frr} would be of order $N^2$ and so the off-diagonal terms
would dominate and there would be instabilities. However, on
$\BS^5$ the distribution of the relative angle $\cos
\vartheta=\BOmega_p\cdot\BOmega_q$ is weighted by a factor
$\sin^4\vartheta$. Hence, near $\vartheta=0$ the average
separation in $\vartheta$ is $\delta\vartheta$ where
\be
\frac{8N}{3\pi}\int_0^{\delta\vartheta}d\vartheta\,\sin^4\vartheta=1
\ee
{\it i.e.\/}, $\delta\vartheta\sim N^{-1/5}$. Hence, the
off diagonal terms in
\eqref{frr} are at most ${\cal O}(N^{2/5})$ and consequently
are subleading at large $N$ and we can ignore them for the purposes
of establishing stability.

To summarize, the issue of stability is determined solely by the
diagonal terms; however, we shall find it necessary to go up to
quartic order in order to settle the issue.

To all orders, we have
\SP{
S+\delta S&=
N\sum_{p=1}^N\Big(|\tfrac1{\sqrt2}\BOmega_p+\delta\Bx_p|^2
+\sigma\theta_p^2\\ &
-\sum_{q=1}^N\log\big((1+\sqrt2\BOmega_p\cdot\delta\Bx_p)(1-\BOmega_p\cdot
\BOmega_q)+|\delta\Bx_p|^2+\theta_p^2-\sqrt2\delta_\perp
\Bx_p\cdot\BOmega_q\big)\Big)\ ,
}
where $\delta_\perp\Bx$ are the variation perpendicular to
$\BOmega_p$, {\it i.e.\/}~tangent to $\BS^5$. We can evaluate
$\delta S$ by replacing the sum over $\BOmega_q$ by an integral
$\int d^5\BOmega$ and use
\EQ{
\sum_q{\cal F}(\BOmega_p\cdot\BOmega_q,\delta_\perp\Bx_p\cdot\BOmega_q)
\longrightarrow \frac{2N}\pi\int_0^\pi d\psi\,\sin^3\psi\,\int_0^\pi
\sin^4\vartheta\,{\cal
  F}(\cos\vartheta,|\delta_\perp\Bx_p|\sin\vartheta
\cos\psi)\ .
}
This gives to quartic order,
\SP{
\delta S=&\frac N3\sum_{p=1}^N\Big((3\sigma-4)\theta_p^2+2
(\BOmega_p\cdot\delta\Bx_p)^2+2\sqrt2(\BOmega_p\cdot\delta\Bx_p)
\big(|\delta\Bx_p|^2+2\theta_p^2\big)\\
&\qquad+|\Bx_p|^4+4|\delta\Bx_p|^2\theta_p^2+
6\theta_p^4\Big)+\cdots\ . }
From this, a careful analysis reveals
that the solution is stable so long as $\sigma>\tfrac43$ as
claimed in the text. Notice that the fluctuations tangent to the
$\BS^5$ are only stable to quartic order.

{\it The low temperature $\BS^5$}

We can prove the stability of this solution using exactly the same
arguments. First of all, the action from Section 3 is
\EQ{
S=\frac{\pi^2\beta R}{g^2}\sum_{p=1}^N|\Bphi_p|^2
-\frac\beta2\sum_{pq=1}^N|\Bphi_{pq}| } and the solution consists
of an $\BS^5$ with $\Bphi_p=r\BOmega_p$, where we have defined the
radius $r=1024\lambda/(945\pi^3R)$. As above only the diagonal
terms play an important r\^ole in the large $N$ limit. Similar
methods give
\SP{
\delta S&=\frac {N\pi^2\lambda\beta}{8R}\sum_{p=1}^N\Big((\BOmega_p\cdot\delta\Bphi_p)^2+
\frac{945\pi^3}{2048}(\BOmega_p\cdot\delta\Bphi_p)
\big((\BOmega_p\cdot\delta\Bphi_p)^2-6|\delta\Bphi_p|^2\big)\\ &-\frac{893025\pi^6}{
33554432}\big(5(\BOmega_p\cdot\delta\Bphi_p)^4+24(\BOmega_p
\cdot\delta\Bphi_p)
|\delta\Bphi_p|^2-24|\delta\Bphi_p|^4\big)\Big)+\cdots\ . } A
careful analysis of this reveals expression proves that the
solution is stable against all fluctuations.

\section{A bestiary of saddles breaking R symmetry}
\label{sec:saddles}

As well as the absolute minima of the action discussed so far, we
can also find various solutions to the effective action that do
not exhibit SO(6) invariance. These solutions break the maximal R
symmetry in certain patterns as we describe below. An important
observation is that at any temperature they have bigger action
than the dominant maximally $SO(6)_R$ symmetric solutions which we
have described so far. Therefore they are all unstable or
metastable. These solutions may be relevant for studies of the
theory at finite chemical potential.

One interesting question raised by the existence of these saddle
points is whether they survive into the strong coupling regime. If
they do, then presumably one should expect to find corresponding (unstable)
supergravity solutions with reduced R symmetry preserved.
Of the solutions that we are about to list, the lower dimensional spheres were
essentially considered in \cite{Berenstein:2005jq}, whereas the
products and fibrations of spheres are new.

As well as the solutions described here, there are also special
two dimensional `Coulomb gas' solutions described in
\cite{Hartnoll:2006pj}. In those solutions only one scalar field
is non-zero. That solution is also different to those listed below in that it uses the
$\theta$ direction in a non-trivial way.

\noindent{\it Low temperatures: $TR \ll 1$}

Within this range of temperature, the $\q$ distribution is
uniform. Therefore there is always an $\BS^1$ part of the eigenvalue
distribution. In the order of increasing action, the solutions we
have found are as follows:

\begin{itemize}
\item{$\BS^1\times \BS^1\times \BS^3$:} The eigenvalue density is
\be\lab{s1s3}
\r(\q,\xb) = \frac{\d(|\vx|-A)\d(|\vy|-\a A)}{8\pi^4\a^3A^4},
\ee
where $\a$ is the ratio of the $\BS^3$ and the $\BS^1$ and the vectors
$\vx$ and $\vy$ span 2D and 4D planes respectively. The equations
of motions for $\vx$ and $\vy$ determine the radius $A$ and the
ratio $\a$ as
\be\lab{r1}
A = \frac{\l C_1}{2\pi^4 TR}\, ,
\ee
with $C_1 = 3.8921\dots$ and $\a = 1.4344\dots$. The action
evaluated on this solution reads
\be\lab{action1}
\frac{1}{N^2}\,S_{\BS^1 \times \BS^1\times \BS^3} = -\frac{\l}{TR} \frac{(1+\a^2)C_1^2}{4\pi^6}\, .
\ee
The difference between the actions of the maximally symmetric $\BS^1
\times \BS^5$ solution and this $\BS^1 \times \BS^1\times \BS^3$ solution is then
\be\lab{diff1}
\frac{TR}{\lambda} \frac{1}{N^2} \Delta S \equiv \frac{TR}{\lambda} \frac{1}{N^2} \left[ S_{\BS^1 \times \BS^1\times
\BS^3}-S_{\BS^1 \times \BS^5}
\right] \approx 0.98 \times 10^{-5} \, .
\ee

\item{$\BS^1\times \BS^2\times \BS^2$:} The eigenvalue density is
$\r(\q,\xb) = \frac{1}{32\pi^3A^4} \d(|\vx|-A)\d(|\vy|-A),$ where
the vectors $\vx$ and $\vy$ span 3D planes. The radius $A$, action
and difference in actions with the $\BS^1 \times \BS^5$ solution are
\bea\lab{r3}
A & = & \frac{2\l(2\sqrt{2}-1)}{15\pi^2 R T } \,, \nonumber \\
\frac{1}{N^2}\,S_{\BS^1\times \BS^2\times \BS^2} & = &
-\frac{\l}{R T}\frac{8}{\pi^2}\le(\frac{2\sqrt{2}-1}{15}\ri)^2\, ,
\nonumber \\
\frac{TR}{\l} \frac{1}{N^2} \Delta S & \approx & 0.103\times 10^{-4} \, .
\eea

\item{$\BS^1\times \BS^1\times \BS^1\times \BS^1$:} The eigenvalue density is
$ \r(\q,\xb) =
\frac{1}{16\pi^4A^3}\d(|\vx|-A)\d(|\vy|-A)\d(|\vz|-A)\,,$
where the vectors $\vx$, $\vy$ and $\vz$ span 2D planes. The
radius $A$, action and difference in actions with the $\BS^1
\times \BS^5$ solution are
\bea\lab{r4}
A & = & \frac{\l C_2}{16\pi^5 TR} \qquad  [C_2=98.7075\dots] \,,
\nonumber \\
\frac{1}{N^2}\, S_{\BS^1\times \BS^1\times \BS^1\times \BS^1} & = &
-\frac{\lambda}{TR}\frac{3}{256\pi^8}C_2^2\,. \nonumber \\
\frac{TR}{\l} \frac{1}{N^2} \Delta S & \approx & 0.209\times 10^{-4}
\, .
\eea

\item{$\BS^1\times \BS^4$:} The eigenvalue density is
$\r(\q,\xb) = \frac{3}{16\pi^3A^4}\d(|\vx|-A)\d(x_6),$ where the
vectors $\vx$ span a 5D plane. We have
\bea\lab{r5}
A & = & \frac{12\l}{35\pi^2 TR} \,, \nonumber \\
\frac{1}{N^2}\,S_{\BS^1 \times \BS^4} & = & -\frac{\l}{R
T}\frac{12^2}{35^2\pi^2}\,, \nonumber \\
\frac{R T}{\l} \Delta S & \approx & 0.144\times 10^{-3} \,.
\eea

\item{$\BS^1\times \BS^1\times \BS^2$:} The eigenvalue density is
$\r(\q,\xb) = \frac{1}{16\pi^3\a^2A^3}\d(|\vx|-A)\d(|\vy|-\a
A)\d(x_6)\,,$ where $\a$ is the ratio of the $\BS^2$ and the $\BS^1$
and the vectors $\vx$ and $\vy$ span 2D and 3D planes
respectively. We have
\bea\lab{r2}
A & = & \frac{\l C_3}{8\pi^3 R T} \, \qquad [C_3 = 5.4034 \dots]
 \,, \nonumber \\
\frac{1}{N^2}\,S_{\BS^1\times \BS^1\times \BS^2} & = & -\frac{\l}{R
T}\frac{(1+\a^2)C_3^2}{64\pi^4}\, \qquad [\a = 1.2404\dots] \,, \nonumber \\
\frac{TR}{\l} \frac{1}{N^2} \Delta S & \approx & 0.165\times 10^{-3} \,.
\eea

\item{$\BS^1\times \BS^3$:} The eigenvalue density is
$\r(\q,\xb) = \frac{1}{4\pi^3A^3}\d(|\vx|-A)\d^2(\vy)\,,$ where
the vectors $\vx$ and $\vy$ span 4D and 2D planes. We have
\bea\lab{r6}
A & = &  \frac{16\l}{15\pi^3 R T} \,, \nonumber \\
\frac{1}{N^2}\,S_{\BS^1 \times \BS^3} & = & -\frac{\l}{R
T}\frac{16^2}{15^2\pi^4}\,, \nonumber \\
\frac{TR}{\l} \frac{1}{N^2} \Delta S & \approx & 0.374\times 10^{-3} \, .
\eea

\item{$\BS^1\times \BS^1\times \BS^1$:} The eigenvalue density is
$ \r(\q,\xb) =
\frac{1}{8\pi^3A^2}\d(|\vx|-A)\d(|\vy|-A)\d^2(\vz)\,,$
where the vectors $\vx$, $\vy$ and $\vz$ span 2D planes. We have
\bea\lab{r9}
A & = & \frac{\l C_4}{8\pi^4 R T} \quad [C_4=18.912\dots] \,,
\nonumber \\
\frac{1}{N^2}\,S_{\BS^1\times \BS^1 \times \BS^1} & = &  -\frac{\l}{R
T}\frac{C_4^2}{32\pi^6}\,, \nonumber \\
\frac{TR}{\l} \frac{1}{N^2} \Delta S & \approx & 0.428\times 10^{-3} \,.
\eea

\item{$\BS^1\times \BS^2$:} The eigenvalue density is
$\r(\q,\xb) = \frac{1}{8\pi^2A^2}\d(|\vx|-A)\d^3(\vy)\,,$ where
the vectors $\vx$ and $\vy$ span 3D planes. We have
\bea\lab{r7}
A & = & \frac{\l}{3\pi^2 R T}\,, \nonumber \\
\frac{1}{N^2}\,S_{\BS^1 \times \BS^2} & = &
-\frac{\l}{TR}\frac{1}{9\pi^2}\,,\nonumber \\
\frac{TR}{\l} \frac{1}{N^2} \Delta S & \approx & 0.796\times 10^{-3}
\,.\nonumber \\
\eea

\item{$\BS^1\times \BS^1$:} The eigenvalue density is
$\r(\q,\xb) = \frac{1}{4\pi^2A}\d(|\vx|-A)\d^4(\vy)\,,$ where the
vectors $\vx$ and $\vy$ span 2D and 4D planes. We have
\bea\lab{r8}
A & = & \frac{\l}{\pi^3 R} \,, \nonumber \\
\frac{1}{N^2}\,S_{\BS^1 \times \BS^1} & = &
-\frac{\l}{TR}\frac{1}{\pi^4}\,, \nonumber \\
\frac{TR}{\l} \frac{1}{N^2} \Delta S & \approx & 0.179\times 10^{-2} \,.
\eea

\end{itemize}

\noindent {\it Intermediate temperatures: $1 \ll TR \ll \l^{-1/2}$}

For temperatures high enough above the deconfinement transition,
we have shown above that the $\q$ distribution is gapped. To
zeroth order in $\l$ it is given by,
\be\lab{qdist} \r(\q) = 4\pi (TR)^3 \sqrt{\q_0^2-\q^2},\qquad
\q_0^2 = \frac{1}{2\pi^2(TR)^3}\,.
\ee
The intermediate temperature joint eigenvalue densities are
obtained from the solutions listed in the previous subsection by
multiplying the eigenvalue densities of the previous subsection by
$2\pi\r(\q)$ and letting the radii $A$ become a function of
$\theta$, which we will denote $r(\theta)$, where
\be\lab{radius}
r(\q) = \frac{C\l}{\sqrt{2}\pi R T}\r(\q)\, ,
\ee
where $C$ is now a constant that will depend on the particular
solution, as we shall see below.

Combining equations \eqref{qdist} and \eqref{radius} we find that
the solutions within this range of temperature have the
ellipsoidal like form:
\be\lab{ellipse}
\frac{\pi^2}{4 C^2 \lambda^2 TR} r^2 + 2\pi^2 (TR)^3 \q^2= 1\,.
\end{equation}
The topology of these solutions will no longer be $\BS^6$ however,
but rather lower dimensional spheres and various singular spaces.

\begin{itemize}

\item The $\BS^1 \times \BS^1\times \BS^3$ solution of the previous section
becomes $\BS^1\times \BS^3$ fibred over an interval. Topologically,
this is a singular space which may be described as an $\BS^5$ in
which a linked $\BS^1$ and $\BS^3$ have been pinched to a point. The
space is described by \eqref{ellipse} together with
\begin{equation}\label{ss1s3}
x_1^2+x_2^2=\frac{1}{\a^2}(x_3^2+\cdots+x_6^2)=r^2\,.
\end{equation}
The coefficient in \eqref{radius} is $C = 0.557 ...$ and as before
the ratio $\a = 1.4344\dots$.

\item The $\BS^1 \times \BS^2\times \BS^2$ solution of the previous section becomes
$\BS^2\times \BS^2$ fibred over an interval. This is a singular space
given by $\BS^5$ in which two linked $\BS^2$s are pinched to a point.
It is described by \eqref{ellipse} together with
\begin{equation}\label{ss2s2}
x_1^2+x_2^2+x_3^2=x_4^2+x_5^2+x_6^2=r^2\,.
\end{equation}
The coefficient $C = \frac{4}{15}(4-\sqrt{2})$.

\item The $\BS^1\times \BS^1\times \BS^1\times \BS^1$
solution of the previous section becomes $\BS^1\times \BS^1\times \BS^1$
fibred over an interval. It is given by \eqref{ellipse} together
with
\begin{equation}\label{ss1s1s1}
x_1^2+x_2^2=x_3^2+x_4^2=x_5^2+x_6^2=r^2\,.
\end{equation}
The coefficient $C=0.5628 \dots$.

\item The $\BS^4\times \BS^1$ solution becomes
a squashed $\BS^5$, given by \eqref{ellipse} together with
\begin{equation}\label{ss4}
x_1^2+\cdots+x_5^2=r^2 \,, x_6 = 0 \,.
\end{equation}
The coefficient $C = \frac{24\sqrt{2}}{35}$.

\item The $\BS^1 \times \BS^1\times \BS^2$ solution becomes
$\BS^1\times \BS^2$ fibred over an interval. Topologically this is an
$\BS^4$ where a linked $\BS^1$ and $\BS^2$ have been pinched to a point.
It is given by \eqref{ellipse} together with
\begin{equation}\label{ss1s2}
x_1^2+x_2^2=\frac{1}{\a^2}(x_3^2+x_4^2+x_5^2)=r^2\,, x_6 = 0 \,.
\end{equation}
The coefficient $C=0.608\dots$, and as before the ratio $\a =
1.2404\dots$.

\item The $\BS^3\times \BS^1$ solution becomes a squashed $\BS^4$,
  given by \eqref{ellipse} together with
\begin{equation}\label{ss3}
x_1^2+\cdots+x_4^2=r^2\,, x_5 = x_6 = 0 \,.
\end{equation}
The coefficient $C = \frac{32\sqrt{2}}{15\pi}$.

\item The $\BS^1\times \BS^1\times \BS^1$ solution becomes
$\BS^1\times \BS^1$ fibred over an interval, which is topologically
and $\BS^3$ with two linked $\BS^1$s pinched to a point. It is given
by \eqref{ellipse} together with
\begin{equation}\label{ss1s1}
x_1^2+x_2^2=x_3^2+x_4^2=r^2\,, x_5 = x_6 = 0 \,.
\end{equation}
The coefficient $C=0.169\dots$.

\item The $\BS^1\times \BS^2$ solution goes over to
a squashed $\BS^3$, given by \eqref{ellipse} together with
\begin{equation}\label{ss2}
x_1^2+x_2^2+x_3^2=r^2\,, x_4 = x_5 = x_6 = 0 \,.
\end{equation}
The coefficient $C = \frac{2\sqrt{2}}{3}$.

\item The $\BS^1\times \BS^1$ solution goes over to a
squashed $\BS^2$, given by \eqref{ellipse}
together with
\begin{equation}\label{ss1}
x_1^2+x_2^2=r^2 \,, x_3 = \cdots = x_6 = 0 \,.
\end{equation}
The coefficient $C = \frac{2\sqrt{2}}{\pi}$.

\end{itemize}

\noindent {\it High Temperatures: $TR\sim \l^{-1/2}$ }

In this range of $T$, recall that the relevant action is
\bea\lab{highact}
S & = & \int d\q d^6x \r(\q, \xb)\le(P |\xb|^2+Q\q^2\ri) \nonumber \\
 & - & \half\int
d\q d\q' d^6x d^6x' \r(\q,\xb)\r(\q',\xb')
\log\le(|\xb-\xb'|^2+(\q-\q')^2\ri)\, .
\eea

As the temperature increases the width of the $\theta$
distribution becomes narrower. Similarly to the $SO(6)$ symmetric
case, we expect a second order phase transition to occur at a
critical temperature, and for $\rho(\q)$ to collapse to a delta function
above that temperature. We will now describe R symmetry breaking
solutions with
\be\lab{highTrho}
\rho(\q) = \d(\q).
\ee

One finds the solutions and evaluates their actions in
the same spirit as in the previous subsections. It
turns out that the order of the actions evaluated on these
solutions stay the same and one finds the following ordering:
\be\lab{order}
S_{\BS^1\times \BS^3}<S_{\BS^2\times \BS^2}<S_{\BS^1\times \BS^1\times
\BS^1}<S_{\BS^4} <S_{\BS^1\times \BS^2}<S_{\BS^3}<S_{\BS^1\times
\BS^1}<S_{\BS^2}<S_{\BS^1}\,.
\ee
The actions have the following values, putting $P=1$ without loss of generality,
respectively: 0.556775,
0.556853, 0.559185, 0.570093, 0.574554, 0.596574, 0.610025,
0.653427, 0.846574. We note that the first three solutions in \eqref{order} are
extremely close to the value of the maximally symmetric solution
$\BS^5$ (0.554907).

\end{document}